\documentclass[10pt,a4paper]{article}
 \usepackage[margin=3cm]{geometry}
\usepackage{graphicx}    
 \usepackage{amsmath,amsfonts}
 \usepackage{color}
 \definecolor{darkgreen}{rgb}{0.05,0.6,0.1}
 \definecolor{darkblue}{rgb}{0.05,0.1,0.6}
 \usepackage{enumerate,subfigure,array}
  \newcolumntype{g}{>{$}c<{$}}

 \newcommand\enquote[1]{``#1''}

 \newcommand\sstyle[1]{{\scriptscriptstyle \text{#1}}}
 \newcommand\sss[2]{#1^\sstyle{#2}}
 
 \def\rrD   {\sss{\rho}{D}} 
 \def\rrN   {\sss{\rho}{N}} 
 \def\rrJ   {\sss{\rho}{J}} 
 \def\rD    {\sss{r}{D}} 
 \def\rN    {\sss{r}{N}} 
 \def\rJ    {\sss{r}{J}} 
 \def\rDaily    {\sss{r}{daily}} 
 \def\sD    {\sss{s}{D}} 
 \def\sN    {\sss{s}{N}} 
 \def\sJ    {\sss{s}{J}} 
 \def\vD    {\sss{v}{D}} 
 \def\vN    {\sss{v}{N}} 
 \def\sigD  {\sss{\sigma}{D}} 
 \def\sigN  {\sss{\sigma}{N}} 
 \def\sigJ  {\sss{\sigma}{J}} 
 \def\xiD   {\sss{\xi}{D}} 
 \def\xiN   {\sss{\xi}{N}} 
 \def\xiJ   {\sss{\xi}{J}} 
 \def\nuD   {\sss{\nu}{D}} 
 \def\nuN   {\sss{\nu}{N}} 
 \def\nuJ   {\sss{\nu}{J}} 
 \def\nuDaily   {\sss{\nu}{daily}} 
 \def\ThetaD{\sss{\Theta}{D}} 
 \def\ThetaN{\sss{\Theta}{N}} 
 \def\ThetaJ{\sss{\Theta}{J}} 

\newcommand{\KK}[2][]{K^{\sstyle{#2} \to \sstyle{#1}}}
\newcommand{\LL}[2][]{L^{\sstyle{#2} \to \sstyle{#1}}}
\newcommand{\hK}[2][]{\widehat{K}^{\sstyle{#2} \to \sstyle{#1}}}

\begin{document}

\title{The fine structure of volatility feedback II:\\ overnight and intra-day effects}
\author{Pierre~Blanc$^{1,2}$%
\footnote{Corresponding author: blancp@cermics.enpc.fr. 
          P.~B. realized most of this work during an internship at Capital Fund Management in 2012.
          He is currently a PhD student at CERMICS (Universit\'e Paris-Est) and benefits from a half-scholarship from Fondation Natixis.} 
 \and R\'emy~Chicheportiche$^{1,3}$%
\footnote{R.C. is now a researcher at Swissquote Bank, 1196 Gland, Switzerland} 
 \and Jean-Philippe~Bouchaud$^1$\\[15pt]
{\normalsize  $^1$ Capital~Fund~Management, 75\,007 Paris, France}\\
{\normalsize  $^2$ Universit\'e Paris-Est, CERMICS, Project team MathRISK ENPC-INRIA-UMLV}\\
{\normalsize  \'Ecole des Ponts, 77\,455 Marne-la-vall\'ee, France}\\
{\normalsize  $^3$ Chaire de finance quantitative, \'Ecole Centrale Paris, 92\,295 Ch\^atenay-Malabry, France}
}
\date{\today}

\maketitle

\begin{abstract}
We decompose, within an ARCH framework, the daily volatility of stocks into overnight and intra-day contributions. 
We find, as perhaps expected, that the overnight and intra-day returns behave completely differently. For example, 
while past intra-day returns affect equally the future intra-day and overnight volatilities, past overnight returns have a weak effect
on future intra-day volatilities (except for the very next one) but impact substantially future overnight volatilities. 
The exogenous component of overnight volatilities is found to be close to zero, which means that the lion's share of 
overnight volatility comes from feedback effects. The residual kurtosis of returns is small for intra-day returns but infinite for 
overnight returns. We provide a plausible interpretation for these findings, and show that our Intra-day/Overnight model 
significantly outperforms the standard ARCH framework based on daily returns for Out-of-Sample predictions.
\end{abstract}


\clearpage
\section{Introduction}

The ARCH (auto-regressive conditional heteroskedastic) framework was introduced in \cite{engle1982autoregressive} 
to account for volatility clustering in financial markets and other economic time series.
It posits that the current relative price change $r_t$ can be written as the product of a 
``volatility'' component $\sigma_t$ and a certain random variable $\xi_t$, of zero mean and unit variance, 
and that the dynamics of the volatility is self-referential in the sense that it depends on the past returns themselves as:
\begin{equation}\label{eq:standard_ARCH}
\sigma_t^2 = s^2  +  \sum_{\tau=1}^q K(\tau) \,      r_{t-\tau}^2 
      \equiv s^2  +  \sum_{\tau=1}^q K(\tau) \, \sigma_{t-\tau}^2 \xi_{t-\tau}^2,
\end{equation}
where $s^2$ is the ``baseline'' volatility level, that would obtain in the absence of any feedback from the past, 
and $K(\tau)$ is a kernel that encodes the strength of the influence of past returns. 
The model is well defined and leads to a stationary time series whenever the feedback is not too strong, 
i.e.\ when $ \sum_{\tau=1}^q K(\tau) < 1$. 
A very popular choice, still very much used both in the academic and professional literature, 
is the so called ``GARCH'' (Generalized ARCH), that corresponds to an exponential kernel, 
$K(\tau)=g e^{-\tau/\tau_\text{p}}$, with $q \to \infty$. 
However, the long-range memory nature of the volatility correlations in financial markets suggests that 
a power-law kernel is more plausible --- a model called ``FIGARCH'', see \cite{bollerslev1994arch,baillie1996fractionally} and below.

Now, the ARCH framework implicitly singles out a {\it time scale}, 
namely the time interval over which the returns $r_t$ are defined. 
For financial applications, this time scale is often chosen to be one day, 
i.e.\ the ARCH model is a model for daily returns, 
defined for example as the relative variation of price between two successive closing prices. 
However, this choice of a day as the unit of time is often a default imposed by the data itself. 
A natural question is to know whether other time scales could also play a role in the volatility feedback mechanism. 
In our companion paper \cite{Chicheportiche2014}, we have studied this question in detail,
focusing on time scales \emph{larger than} (or equal to) the day. 
We have in fact calibrated the most general model, called ``QARCH'' \cite{sentana1995quadratic}, 
that expresses the squared volatility as a quadratic form of past returns, i.e.\ 
with a two-lags kernel $K(\tau,\tau')r_{t-\tau}r_{t-\tau'}$ instead of the ``diagonal'' regression \eqref{eq:standard_ARCH}. 
This encompasses all models where returns defined over arbitrary time intervals could play a role, 
as well as (realized) correlations between those --- 
see \cite{Chicheportiche2014} and references therein for more precise statements, and \cite{muller1997volatilities,zumbach2001heterogeneous,borland2005multi,shapira2011hidden} 
for earlier contributions along these lines. 

The main conclusion of our companion paper \cite{Chicheportiche2014} is that 
while other time scales play a statistically significant role in the feedback process
(interplay between $r_{t-\tau}$ and $r_{t-\tau'}$ resulting in non-zero off-diagonal elements $K(\tau,\tau')$), 
the {\it dominant effect} for daily returns is indeed associated with past daily returns. 
In a first approximation, a FIGARCH model based on daily returns, 
with an exponentially truncated power-law kernel $K(\tau)=g \tau^{-\alpha} e^{-\tau/\tau_\text{p}}$, 
provides a good model for stock returns, with $\alpha \approx 1.1$ and $\tau_\text{p} \approx 50$ days. 
This immediately begs the question: if returns on time scales larger than a day appear to be of lesser importance,%
\footnote{Note a possible source of confusion here since a FIGARCH model obviously involves many time scales. 
We need to clearly distinguish time {\it lags}, as they appear in the kernel $K(\tau)$, from time scales, 
that enter in the definition of the returns themselves.}
what about returns on time scales smaller than a day? 
For one thing, a trading day is naturally decomposed into trading hours, that define an `Open to Close' (or `intra-day') return, 
and hours where the market is closed but news accumulates and impacts the price at the opening auction, 
contributing to the `Close to Open' (or `overnight') return. 
One may expect that the price dynamics is very different in the two cases, for several reasons. 
One is that many company announcements are made overnight, that can significantly impact the price. 
The profile of market participants is also quite different in the two cases: 
while low-frequency participants might choose to execute large volumes during the auction, 
higher-frequency participants and market-makers are mostly active intra-day. 
In any case, it seems reasonable to distinguish two volatility contributions, 
one coming from intra-day trading, the second one from overnight activity. 
Similarly, the feedback of past returns should also be disentangled into an intra-day contribution and an overnight contribution. 
The calibration of an ARCH-like model that distinguishes between intra-day and overnight returns is the aim of the present paper, 
and is the content of Section~\ref{DON_returns}. 
We have in fact investigated the role of higher frequency returns as well. 
For the sake of clarity we will not present this study here, but rather summarize briefly our findings on this point in the conclusion.

The salient conclusions of the present paper are that the intra-day and overnight dynamics are indeed completely different 
--- for example, while the intra-day (Open-to-Close) returns impact both the future intra-day and overnight volatilities in a slowly decaying manner, 
overnight (Close-to-Open) returns essentially impact the next intra-day but very little the following ones. 
However, overnight returns have themselves a slowly decaying impact on future overnights. 
Another notable difference is the statistics of the residual factor $\xi_t$, which is {\it nearly Gaussian} for intra-day returns, 
but has an {\it infinite kurtosis} for overnight returns. 
We discuss further the scope of our results in the conclusion Section~\ref{discussion}, 
and relegate several more technical details to appendices.

\section{The dynamics of Close-to-Open and Open-to-Close stock returns and volatilities}\label{DON_returns}

Although the decomposition of the daily (Close-to-Close) returns into their intra-day and overnight components 
seems obvious and intuitive, very few attempts have actually been made to model them jointly (see \cite{gallo2001modelling,tsiakas2008overnight}). 
In fact, some studies even discard overnight returns altogether. 
In the present section, we define and calibrate an ARCH model that explicitly treats these two contributions separately. 
We however first need to introduce some precise definitions of the objects that we want to model.

\subsection{Definitions, time-line and basic statistics}

We consider equidistant time stamps $t$ with $\Delta t=1$ day. 
Every day, the prices of traded stocks are quoted from the opening to the closing hour,
but we only keep track of the first and last traded prices. 
For every stock name $a$, $O_t^a$ is the open price and $C_t^a$ the close price at date $t$.
(In the following, we drop the index $a$ when it is not explicitly needed).
We introduce the following definitions of the geometric returns, volatilities, and residuals:
\begin{subequations}\label{eq:def_ret}
\begin{align}
\label{eq:def_oc} \text{Intra-day return: }   &                  \rD_t = \ln({C_t}/{O_t})      \phantom{_{-1}} &\equiv \sigD_t  \xiD_t    \\
\label{eq:def_co} \text{Overnight return: }  &                  \rN_t = \ln({O_t}/{C_{t-1}})  \phantom{     } &\equiv \sigN_t  \xiN_t \\
\label{eq:def_cc} \text{Daily return: }      & \sss{r}{\phantom{D}}_t = \ln({C_t}/{C_{t-1}}) = \rD_t + \rN_t  &\equiv \sigma_t \xi_t.
\end{align}
\end{subequations}
The following time-line illustrates the definition of the three types of return:
\begin{equation}\label{eq:timeline}
\ldots \longrightarrow 			          C_{t-1} \quad\left|\quad
								\underbrace{\underbrace{\xrightarrow{\text{Night} \ t}}_{\textcolor{blue}{\rN_t}} 
									\quad O_t \quad 
								\underbrace{\xrightarrow{\quad \text{Day} \ t \quad}}_{\textcolor{green}{\rD_t}}}_{r_t}
									\quad C_t \quad\right|\quad
								\underbrace{\xrightarrow{\text{Night} \ t+1}}_{\textcolor{blue}{\rN_{t+1}}}
                                    \quad  O_{t+1}
\longrightarrow\ldots
\end{equation}
To facilitate the reading of our tables and figures, intra-day returns are associated with the green color (or light gray)
and overnight returns with blue (or dark gray). 
\begin{figure}
 \center
		\includegraphics[width=\textwidth,height=.3\textwidth,trim= 0 75 0  0,clip]{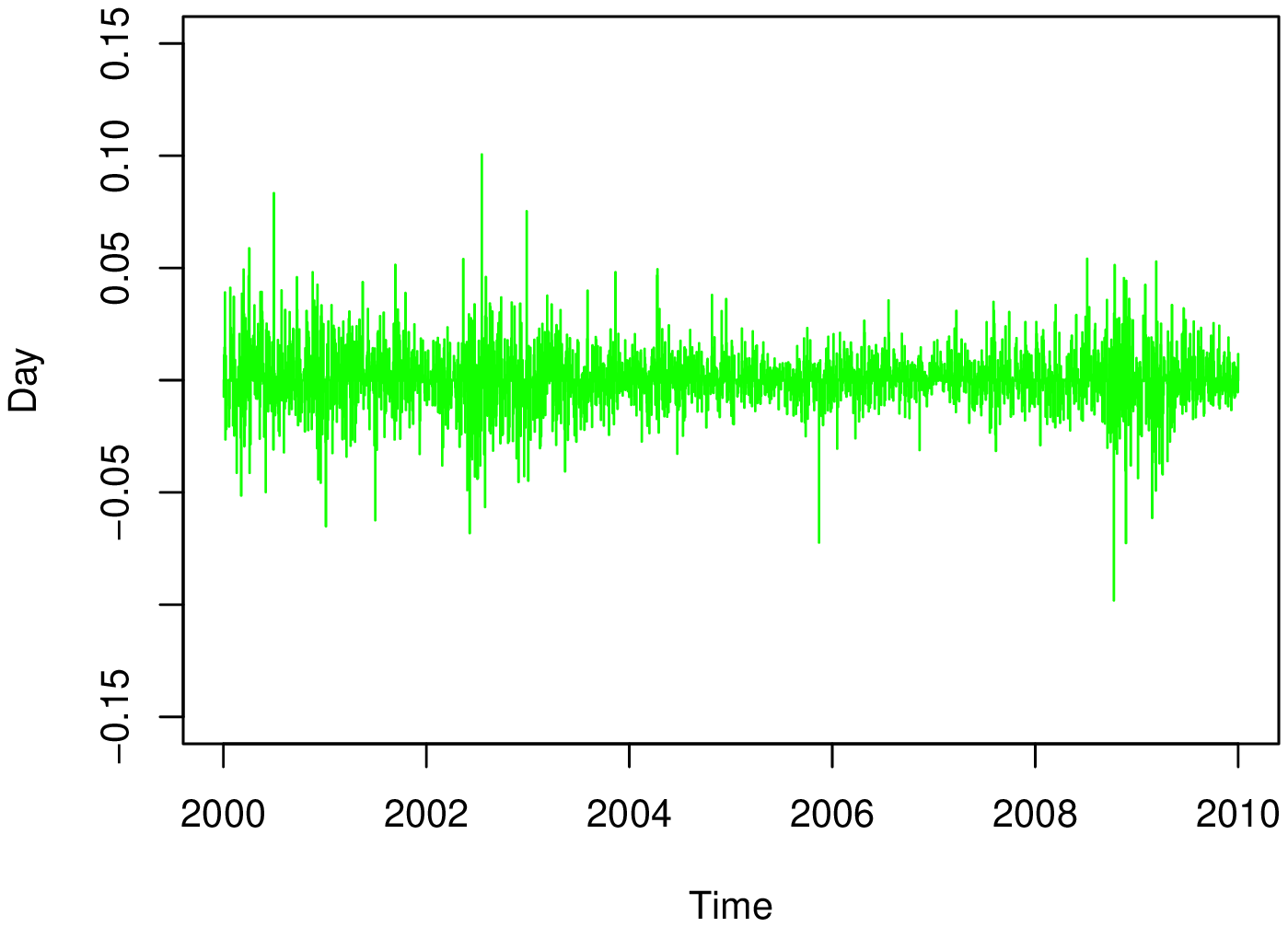}
		\includegraphics[width=\textwidth,height=.3\textwidth,trim= 0  0 0 55,clip]{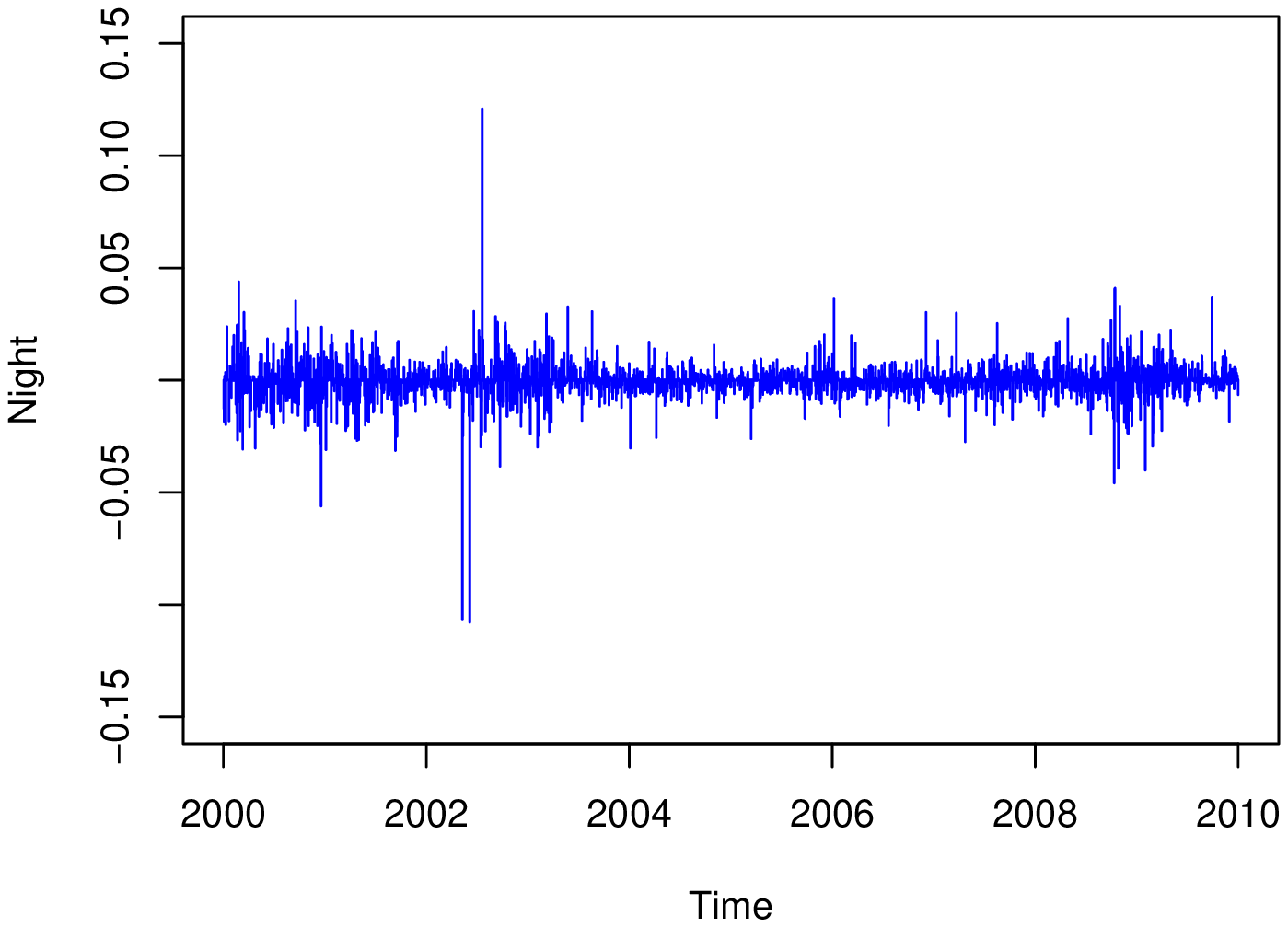}
		\caption{Example of a historical time series of stock day returns (top) and overnight returns (bottom).}
        \label{graph:empirical_ON_ret_serie} 
\end{figure}

Before introducing any model, we discuss the qualitative statistical differences 
in the series of Open-Close returns $\rD_t$ and Close-Open returns $\rN_t$.
First, one can look at Fig.~\ref{graph:empirical_ON_ret_serie} for a visual impression of the difference: 
while the intra-day volatility is higher than the overnight volatility, the relative importance of \enquote{surprises} 
(i.e.\ large positive or negative jumps) is larger for overnight returns. 
This is confirmed by the numerical values provided in Tab.~\ref{tab:descr_stat} for the volatility, 
skewness and kurtosis of the two time  series $\rD_t$ and $\rN_t$. 
\begin{table}[h]
\[\begin{array}{c||cccc}
    \sstyle{J} &           \left[ \langle  \rJ    \rangle \right]   &
               \sigJ=\sqrt{\left[ \langle {\rJ}^2 \rangle \right]}  &
                           \left[ \langle {\rJ}^3 \rangle / \langle {\rJ}^2 \rangle^{\frac 3 2} \right]   &
                           \left[ \langle {\rJ}^4 \rangle / \langle {\rJ}^2 \rangle^2           \right]   \\\hline\hline
    \sstyle{D} & \phantom{-}1.2 \ 10^{-4}& 0.022& -0.12& 12.9\\
    \sstyle{N} &           -1.0 \ 10^{-4}& 0.013& -1.5 & 62.6
\end{array}\]
\caption{Distributional properties of intra-Day and overNight returns (first four empirical moments).
         $\langle \cdot \rangle$  means average over all dates, and $ \left[ \cdot \right]$ average over all stocks.
		 }
\label{tab:descr_stat}
\end{table}

It is also visible on Fig.~\ref{graph:empirical_ON_ret_serie} that periods of high volatilities are common to the two series: 
two minor ones can be observed in the middle of year 2000 and at the beginning of year 2009, 
and an important one in the middle of year 2002.

An important quantity is the correlation between intra-day and overnight returns, 
which can be measured either as $\left[\langle \rN_t \rD_t \rangle \right]/\sigN \sigD$ (overnight leading intra-day)
or as $\left[\langle \rD_t \rN_{t+1} \rangle \right]/\sigN \sigD$ (intra-day leading overnight). 
The  statistical reversion revealed by {the measured values of the above correlation coefficients ($-0.021$ and $-0.009$, respectively)} 
is slight enough (compared to the amplitude and reach of the feedback effect) to justify the assumption of i.i.d.\ residuals.
If there were no linear correlations between intra-day and overnight returns, the squared volatilities would be exactly additive, 
i.e.\ $\sigma_t^2 \equiv {\sigD_t}^2 + {\sigN_t}^2$. Deviations from this simple addition of variance rule are below $2 \%$.

\subsection{The model}

The standard ARCH model recalled in the introduction, Eq.~\eqref{eq:standard_ARCH}, can be rewritten identically as:
\begin{equation}
\label{eq:standard_ARCH_bis}
\sigma_t^2 \equiv  s^2  +   \sum_{\tau=1}^q K(\tau) {\rN_{t-\tau}}^2 
                        +   \sum_{\tau=1}^q K(\tau) {\rD_{t-\tau}}^2 
                        + 2 \sum_{\tau=1}^q K(\tau) \rN_{t-\tau}\rD_{t-\tau},
\end{equation}
meaning that there is a unique kernel $K(\tau)$ describing the feedback of past intra-day and overnight returns on the current volatility level.

If however one believes that these returns are of fundamentally different nature, one should expand the model in two directions: 
first, the two volatilities ${\sigD}^2$ and ${\sigN}^2$ should have separate dynamics. 
Second, the kernels describing the feedback of past intra-day and overnight returns should {\it a priori} be different. 
This suggests to write the following generalized model for the intra-day volatility:
\begin{align}
\label{eq:volD_model}
{\sigD_t}^2 =  {\sD}^2 	&+  \sum_{\tau=1}^\infty \LL[D]{D} (\tau)    \rD_{t-\tau}     \phantom{(+1}
                         +  \sum_{\tau=1}^\infty \KK[D]{DD}(\tau)   {\rD_{t-\tau}}^2  \phantom{(+1}   
                         + 2\sum_{\tau=1}^\infty \KK[D]{ND}(\tau)    \rD_{t-\tau} \rN_{t-\tau}        	\\
\nonumber	            &+  \sum_{\tau=0}^\infty \LL[D]{N} (\tau+1)  \rN_{t-\tau}            
                         +  \sum_{\tau=0}^\infty \KK[D]{NN}(\tau+1) {\rN_{t-\tau}}^2    
                         + 2\sum_{\tau=0}^\infty \KK[D]{DN}(\tau+1)  \rD_{t-\tau-1} \rN_{t-\tau},
\end{align}
where we have added the possibility of a ``leverage effect'', i.e.\  terms linear in past returns 
that can describe an asymmetry in the impact of negative and positive returns on the volatility.
The notation used is, we hope, explicit: for example 
$\KK[D]{DD}$ describes the influence of squared intra-Day past returns on the current intra-Day volatility. 
Note that the mixed effect of intra-Day and overNight returns requires two distinct kernels, 
$\KK[D]{DN}$ and $\KK[D]{ND}$, depending on which comes first in time. 
Finally, the time-line shown above explains why the $\tau$ index starts at $\tau=1$ for past intra-day returns, 
but at $\tau=0$ for past overnight returns.
We posit a similar expression for the overnight volatility:
\begin{align}
\label{eq:volN_model}{
\sigN_t}^2 =  {\sN}^2 	&+  \sum_{\tau=1}^\infty \LL[N]{N} (\tau)  \rN_{t-\tau}  
                         +  \sum_{\tau=1}^\infty \KK[N]{NN}(\tau) {\rN_{t-\tau}}^2
                         + 2\sum_{\tau=1}^\infty \KK[N]{ND}(\tau)  \rD_{t-\tau} \rN_{t-\tau}        	\\
\nonumber	            &+  \sum_{\tau=1}^\infty \LL[N]{D} (\tau)  \rD_{t-\tau}            
                         +  \sum_{\tau=1}^\infty \KK[N]{DD}(\tau) {\rD_{t-\tau}}^2    
                         + 2\sum_{\tau=1}^\infty \KK[N]{DN}(\tau)  \rD_{t-\tau-1} \rN_{t-\tau}.
\end{align}
The model is therefore fully characterized by 
two base-line volatilities $\sD, \sN$, 
four leverage (linear) kernels $\LL[J]{J'}$, 
eight quadratic kernels $\KK[J]{J'J''}$, and 
the statistics of the two residual noises $\xiD, \xiN$ needed to define the returns, as $\rJ=\sigJ \xiJ$. 
We derive in Appendix~\ref{appendice:nneg_vol} conditions on the coefficients of the model 
under which the two volatility processes remain positive at all times. 
The model as it stands has a large number of parameters; 
in order to ease the calibration process and gain in stability, 
we in fact choose to parameterize the $\tau$ dependence of the different kernels with some simple functions, 
namely an exponentially truncated power-law for $\KK[J]{J'J''}$ and a simple exponential for $\LL[J]{J'}$:
\begin{equation}\label{eq:parametricKernels}
K(\tau) = g_\text{p} \tau^{-\alpha} \,\exp\!\left(-\omega_\text{p}\, \tau\right); \qquad 
L(\tau) = g_\text{e} \,\exp\!\left(-\omega_\text{e} \, \tau\right).
\end{equation}
The choice of these functions is not arbitrary, but is suggested by a preliminary calibration of the model using a generalized method of moments (GMM),
as explained in the companion paper, see Appendix~C.2 in Ref.~\cite{Chicheportiche2014}.

As far as the residuals $\xiD_t,\xiN_t$ are concerned, we assume them to be i.i.d.\ centered Student variables 
of unit variance with respectively $\nuD > 2$ and $\nuN > 2$ degrees of freedom. 
Contrarily to many previous studies, we prefer to be agnostic about the kurtosis of the residuals 
rather than imposing a priori Gaussian residuals. 
It has been shown that while the ARCH feedback effect accounts for volatility clustering 
and for some positive kurtosis in the returns, 
this effect alone is not sufficient to explain the observed heavy tails in the return distribution 
(see for example \cite{Chicheportiche2014}). 
These tails come from true `surprises' (often called jumps), 
that cannot be anticipated by the predictable part of the volatility, 
and that can indeed be described by a Student (power-law) distribution of the residuals.

\subsection{Dataset}\label{subsection:data} 

\begin{figure}
	\centering
		\includegraphics[scale=0.6]{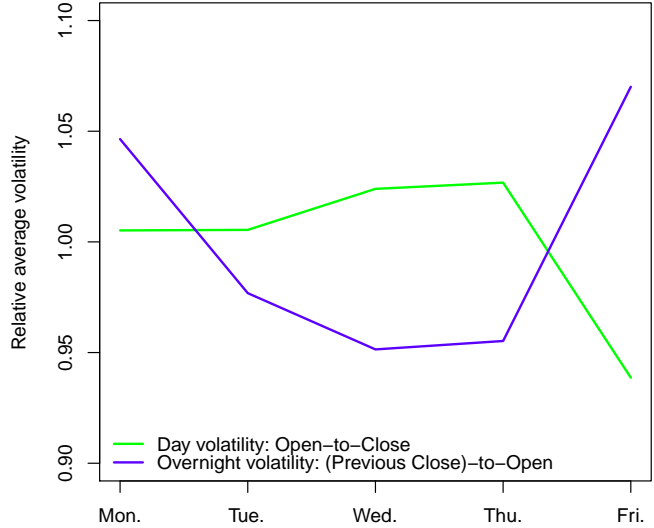}
		\caption{Normalized weekly seasonality of the volatility.
                 The overnight volatility is that of the previous night 
                 (i.e.\ the volatility of the weekend for Monday and that of Thursday night for Friday).
        }
        \label{graph:weekly_seas} 
\end{figure}

The dataset used to calibrate the model is exactly the same as in our companion paper \cite{Chicheportiche2014}. 
It is composed of US stock prices (four points every day: Open, Close, High and Low) 
for $N=280$ stocks present permanently
in the S\&P-500 index from Jan.~2000 to Dec.~2009 ($T=2515$ days). 
For every stock $a$, the daily returns ($r_t^a$), intra-day returns (${\rD_t}^a$) and overnight returns (${\rN_t}^a$)
 defined in Eq.~\eqref{eq:def_ret} are computed using only Open and Close prices at every date $t$.
In order to improve the statistical significance of our results, 
we consider the pool of stocks as a statistical ensemble over which we can average.
This means that we assume a universal dynamics for the stocks, a reasonable assumption as we discuss in Appendix~\ref{appendix:univ_DON}.

Bare stock returns are ``polluted'' by several obvious and predictable events
associated with the life of the company, such as quarterly announcements.
They also reflect low-frequency human activity, such as a weekly cyclical pattern of the volatility, 
which is interesting in itself (see Fig.~\ref{graph:weekly_seas}).
These are of course real effects, but the ARCH family of models we investigate here 
rather focuses on the endogenous self-dynamics {\it on top of} such seasonal patterns. 
For example, the quarterly announcement dates are responsible for returns of typically much larger magnitude 
(approximately three times larger on average for daily returns)
that have a very limited feedback in future volatility.

We therefore want to remove all obvious seasonal effects from the dataset, 
and go through the following additional steps of data treatment before estimating the model.
  For every stock $a$, the average over time is denoted           $ \langle r_{\cdot}^a \rangle := \frac{1}{T} \sum_t r_t^a $,
and for each date $t$, the cross-sectional average over stocks is $[        r_t^{\cdot}       ] := \frac{1}{N} \sum_a {r_t^a} $.
All the following normalizations apply both (and separately) for intra-day returns and overnight returns. 
\begin{itemize}
\item The returns series are first centered around their temporal average: 
       $
        r_t^a \leftarrow r_t^a - \langle r_{\cdot}^a \rangle.
       $
       In fact, the returns are already nearly empirically centered, since the temporal average is small, 
       see Tab.~\ref{tab:descr_stat} above.

\item We then divide the returns by the cross-sectional dispersion:
      \[
        r_t^a \leftarrow r_t^a / \sqrt{\left[ {r_t^{\cdot\neq a}}^2\right]}.
      \]
      This normalization%
      \footnote{
      If the element $a$ is not excluded in the average, the tails of the returns are artificially cut-off:
      when $|r_t^a| \to \infty$, ${|r_t^a|}/{\sqrt{[ {r_{t}^{\cdot}}^2 ]}}$ is capped at $\sqrt N < \infty$.
      }
      removes the historical low-frequency patterns of the volatility, 
      for example the weekly pattern  discussed above (Fig.~\ref{graph:weekly_seas}). 
      In order to predict the ``real'' volatility with the model, one must however re-integrate these patterns back into the $\sigJ$'s.

\item Finally, we normalize stock by stock all the returns by their historical standard deviation: 
      for all stock $a$, for all $t$, 
      \[
        r_t^a \leftarrow r_t^a / \sqrt{\langle {{r_{\cdot}}^{a}}^2 \rangle},
      \]
      imposing $\langle {{{\rD_{\cdot}}^a}}^2 \rangle = \langle {{{\rN_{\cdot}}^a}}^2 \rangle  \equiv 1$.
\end{itemize}
This data treatment allows to consider that the residual volatility of the returns series is
independent of the effects we do not aim at modeling, and that the series of all stocks can reasonably 
be assumed to be homogeneous (i.e.\ identically distributed), both across stocks and across time. 
This is necessary in order to calibrate a model that is translational-invariant in time 
(i.e.\ only the time {\it lag} $\tau$ enters Eqs.~(\ref{eq:volD_model},\ref{eq:volN_model})), 
and also to enlarge the calibration dataset by averaging the results over all the stocks in the pool --- see the discussion 
in Appendix~\ref{appendix:univ_DON}.

\subsection{Model estimation}\label{subsection:estimation} 

Assuming that the distribution of the residuals is a Student law, 
the log-likelihood per point of the model can be written as ($\sstyle{J}=\sstyle{D},\sstyle{N}$):
\begin{align}\label{eq:log_likelihood_day}
          \mathcal{L} & (\ThetaJ , \nuJ | \{\rD , \rN\}) =  
                            \frac {\nuJ}  2 \ln\!\Big| (\nuJ-2)\,  \sigma_t^2(\ThetaJ)            \Big|
                       -    \frac{\nuJ+1} 2 \ln\!\Big| (\nuJ-2) \sigma_t^2(\ThetaJ) \ + \ {\rJ_t}^2 \Big|,
\end{align}
where ${\sigJ_t}^2 = \sigma_t^2(\ThetaJ| \{\rD , \rN\})$ are defined in Eqs.~(\ref{eq:volD_model},\ref{eq:volN_model}),
$\nuJ$ are the degrees of freedom of the Student residuals, and
$\ThetaJ$ denote generically the sets of volatility feedback parameters.

Conditionally on the dataset, we maximize numerically the likelihood of the model, averaged over all dates and all stocks. 

\paragraph{Calibration methodology:}

As mentioned above, we in fact choose to parameterize the feedback kernels as suggested by the results of the method of moments, 
i.e.\ exponentially truncated power-laws for $K$'s and simple exponentials for $L$'s. 
Imposing these simple functional forms allows us to gain stability and readability of the results.  
However, the functional dependence of the likelihood on the kernel parameters is not guaranteed to be globally concave, 
as is the case when it is maximized \enquote{freely}, 
i.e.\ with respect to all individual kernel coefficients $K(\tau)$ and $L(\tau)$, with $1 \leq \tau \leq  q_{\text{free}}$. 
This is why we use a three-step approach:
\begin{enumerate}
\item   A first set of kernel estimates is obtained by the Generalized Method of Moments (GMM), see \cite{Chicheportiche2014}, 
        and serves as a starting point for the optimization algorithm.

\item 	We then run a Maximum Likelihood Estimation (MLE) of the unconstrained kernels based on Eq.~(\ref{eq:log_likelihood_day}), over 
        $6 \times q_{\text{free}}$ parameters for both $\ThetaD$ and $\ThetaN$, with
        a moderate value of maximum lag $q_{\text{free}}=63 \simeq $ three months.
		Taking as a starting point the coefficients of step 1 and maximizing with a gradient descent,
		we obtain a second set of (short-range) kernels.

\item   Finally, we perform a MLE estimation of the parametrically constrained kernels 
        with the functional forms~\eqref{eq:parametricKernels} for $K$'s and $L$'s, 
        which only involves $4 \times 3 + 2 \times 2$ parameters in every set $\ThetaD$ and $\ThetaN$, 
        with now a large value of the maximum lag  $q_{\text{constr}}=512 \simeq $ two years.
        Taking as a starting point the functional fits of the kernels obtained at step 2, and maximizing with a gradient descent,
	    we obtain our final set of model coefficients, shown in Tabs.~\ref{day_coefs},\ref{night_coefs}.
\end{enumerate}

Thanks to step~2, the starting point of step~3 is close enough to the global maximum for the likelihood to be locally concave,
and the gradient descent algorithm converges in a few steps. The Hessian matrix of the likelihood is evaluated 
at the maximum to check that the dependency on all coefficients is indeed concave.

The numerical maximization of the likelihood is thus made on $2$ or $3$ parameters per kernel, 
independently of the chosen maximum lag $q_{\text{constr}}$, that can thus be arbitrarily large with little additional computation cost.

Finally, the degrees of freedom $\nu$ of the Student residuals are determined using two separate one-dimensional likelihood maximizations
(one for $q = q_{\text{free}}$ and one for $q = q_{\text{constr}}$) 
and then included as an additional parameter in the MLE of the third step of the calibration. 
Note that $\nu$ does not vary significantly at the third step, 
which means that the estimation of the volatility parameters at the second step can indeed be done independently from that of $\nu$.

This somewhat sophisticated calibration method was tested on simulated data, obtaining very good results.

\paragraph{The special case $s^2 = 0$:}

We ran the above calibration protocol on intra-day and overnight volatilities separately.

For the overnight model, this led to a slightly negative value of the baseline volatility ${\sN}^2$ (statistically compatible with zero).
But of course negative values of $s^2$ are excluded for a stable and positive volatility process. 
For overnight volatility only, we thus add a step to the calibration protocol, 
which includes the constraint ${\sN}^2=0$ in the estimation of $\KK[N]{DD}$ and $\KK[N]{NN}$ 
(which are the two main contributors to the value of the baseline volatility). 
For simplicity, we consider here that $\langle {\sigN}^2 \rangle = \langle {\rD}^2 \rangle = \langle {\rN}^2 \rangle = 1$. 
We take the results of the preceding calibration as a starting point and freeze all the kernels but $\KK[N]{DD}$ and $\KK[N]{NN}$, 
expressed (in this section only) as follows:
\begin{equation}\label{K_s2=0}
        \KK[N]{DD}(\tau) =          g \, \tau^{-\alpha_1} \,\exp\!\left(-\omega_1 \tau \right),
 \qquad \KK[N]{NN}(\tau) = \gamma \ g \, \tau^{-\alpha_2} \,\exp\!\left(-\omega_2 \tau \right),
\end{equation}
where $g = g(\gamma,\alpha_1,\omega_1,\alpha_2,\omega_2)$ is fixed by the constraint ${\sN}^2=0$:
\begin{equation}\label{g_s2=0}
g(\gamma,\alpha_1,\omega_1,\alpha_2,\omega_2) = \frac{1 - c}{h(\alpha_1,\omega_1) + \gamma \ h(\alpha_2,\omega_2)} ; 
\qquad h(\alpha,\omega) = \sum_{\tau=1}^q \tau^{-\alpha} \exp\!\left(-\omega \tau \right) ,
\end{equation}
with $\gamma > 0$ the ratio of the two initial amplitudes, 
and $c$ the (low) contribution of the fixed `cross' kernels $\KK[N]{ND}$ and $\KK[N]{DN}$ to ${\sN}^2$. 
We then maximize the likelihood of the model with respect to the five parameters $\gamma, \alpha_1, \omega_1, \alpha_2$ and $\omega_2$, 
for which a gradient vector and a Hessian matrix of dimension $5$ can be deduced from equations (\ref{K_s2=0}) and (\ref{g_s2=0}). 
The coefficients and confidence intervals of the kernels $\KK[N]{DD}$ and $\KK[N]{NN}$ are replaced in Sect.~\ref{section:kernels}  
by the results of this final step, along with the corresponding value of the overnight baseline volatility, 
${\sN}^2=0$ in Sect.~\ref{section:baseline}.

For intra-day volatility instead, the results are given just below, in Sect.~\ref{section:kernels}.

\section{Intra-day vs.\ overnight: results and discussions}\label{section:DON_empirical}

The calibration of our generalized ARCH framework should determine three families of parameters: 
the feedback kernels $K$ and $L$, the statistics of the residuals $\xi$ and finally the ``baseline volatilities'' $s^2$. 
We discuss these three families in turn in the following sections.

\subsection{The feedback kernels}\label{section:kernels} 

\begin{figure}[p]
	\centering
    \subfigure[Quadratic $K(\tau)$ kernels in log-log scale. For `DN$\to$N', the absolute value of the (negative) kernel is plotted.]{
    \includegraphics[scale=0.5]{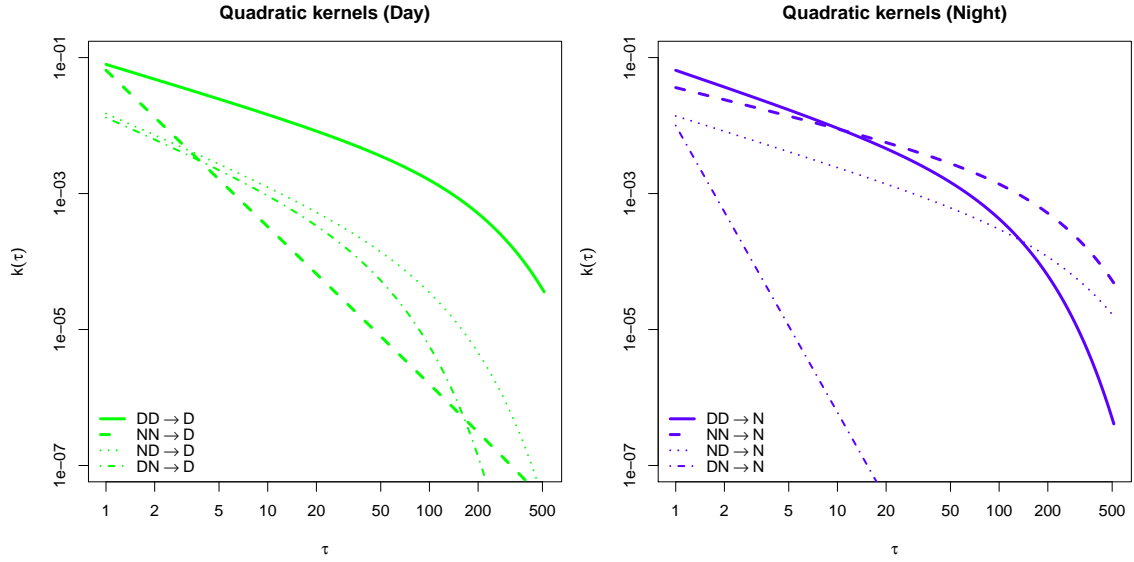}
	\label{graph:Quadratic_kernels_DON_q512}
    }
	\subfigure[Linear $L(\tau)$ kernels in lin-log scale. All leverage kernels are negative, so $-L(\tau)$ are plotted.]{
    \includegraphics[scale=0.6]{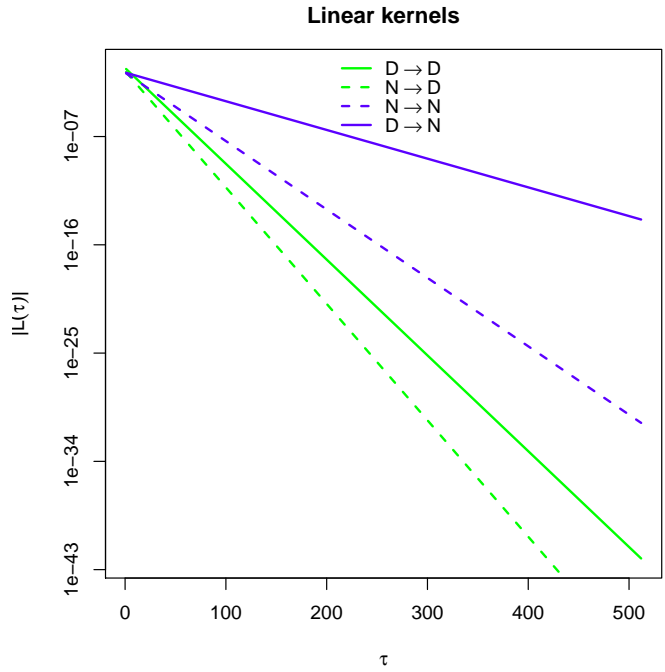}
	\label{graph:Leverage_kernels_DON_q512} 
    }
	\caption{Estimated kernels, impacting intra-Days in (light) green, overNights in (dark) blue.}
    \label{fig:estimated_kernels}
\end{figure}

In this section, we give the results of the ML estimation of the regression kernels for a maximum lag $q=512$:
estimates of the parameters are reported in Tabs.~\ref{day_coefs},\ref{night_coefs}, 
and the resulting kernels are shown in Fig.~\ref{fig:estimated_kernels}. 

\begin{table}[h]
\center
\begin{tabular}{|g||g|g|g||g|g|} \hline
	\text{Kernels}	& 	\multicolumn{3}{g||}{K(\tau)=g_\text{p} \, \tau^{-\alpha} \, e^{-\omega_\text{p} \, \tau}} 											& 	\multicolumn{2}{g||}{L(\tau)=g_\text{e} \, e^{-\omega_\text{e}\, \tau}}  				\\\hline
					& 	g_\text{p} \times 10^2					& \alpha 									& \omega_\text{p} \times 10^2 					&	g_\text{e} \times 10^2					& \omega_\text{e} \times 10^2					\\\hline\hline
  \KK[D]{DD} 		& 	\textcolor{darkgreen}{7.99 \pm 0.06} 	& \textcolor{darkgreen}{0.71 \pm 0.003}		& \textcolor{darkgreen}{0.64 \pm 0.02}			&	--										& --											\\
  \KK[D]{NN} 		& 	\textcolor{darkgreen}{6.53 \pm 0.22} 	& \textcolor{darkgreen}{2.30 \pm 0.07}		& \textcolor{darkgreen}{0.04 \pm 0.97}			&	--										& --											\\
  \KK[D]{ND} 		& 	\textcolor{darkgreen}{1.52 \pm 0.17}	& \textcolor{darkgreen}{1.03 \pm 0.11}		& \textcolor{darkgreen}{1.3 \pm 1.2}			&	--	    								& --											\\
  \KK[D]{DN} 		& 	\textcolor{darkgreen}{1.35 \pm 0.22} 	& \textcolor{darkgreen}{1.03 \pm 0.17}		& \textcolor{darkgreen}{3.0 \pm 4.6}			&	--										& --											\\
  \LL[D]{D} 			& 	--										& --		    							& --											&	\textcolor{darkgreen}{-4.97 \pm 0.25}	& \textcolor{darkgreen}{18.3 \pm 1.3}			\\
  \LL[D]{N} 			& 	--										& --		   								& --											&	\textcolor{darkgreen}{-2.83 \pm 0.30}	& \textcolor{darkgreen}{22.3 \pm 2.5}			\\\hline
\end{tabular}
\caption{Day volatility: estimated kernel parameters for $K$'s and $L$'s, with their asymptotic confidence intervals of level $95 \%$,
                         as computed using the Fisher Information matrix with the Gaussian quantile ($1.98$).}

\label{day_coefs} 
\center
\begin{tabular}{|g||g|g|g||g|g|} \hline
	\text{Kernels}	& 	\multicolumn{3}{g||}{K(\tau)=g_\text{p} \, \tau^{-\alpha} \, e^{-\omega_\text{p}\, \tau}}     							& 	\multicolumn{2}{g||}{L(\tau)=g_\text{e} \, e^{-\omega_\text{e}\, \tau}}  		\\\hline
					& 	g_\text{p} \times 10^2 				& \alpha 								& \omega_\text{p} \times 10^2 				&	g_\text{e} \times 10^2  			& \omega_\text{e} \times 10^2 			    \\\hline\hline
  \KK[N]{DD}			& 	\textcolor{blue}{6.59 \pm 0.33} 	& \textcolor{blue}{0.80 \pm 0.02}		& \textcolor{blue}{1.4 \pm 0.4}				&	--									& --							    		\\
  \KK[N]{NN}			& 	\textcolor{blue}{3.64 \pm 0.17} 	& \textcolor{blue}{0.58 \pm 0.01}		& \textcolor{blue}{0.58 \pm 0.04}			&	--									& --						        		\\
  \KK[N]{ND}   		& 	\textcolor{blue}{1.39 \pm 0.11}		& \textcolor{blue}{0.74 \pm 0.03}		& \textcolor{blue}{0.42 \pm 0.12}			&	--   								& --						        		\\
  \KK[N]{DN}   		& 	\textcolor{blue}{-1.00 \pm 0.29} 	& \textcolor{blue}{4.22 \pm 2.44}		& \textcolor{blue}{0.02 \pm 23}				&	--									& --				  						\\
  \LL[N]{D} 			& 	--									& --	    							& --										&	\textcolor{blue}{-2.09 \pm 0.05}	& \textcolor{blue}{5.5 \pm 0.2}				\\
  \LL[N]{N}  		& 	--									& --	    							& --						    			&	\textcolor{blue}{-2.03 \pm 0.20}	& \textcolor{blue}{13.1 \pm 2.2}			\\\hline
\end{tabular}
\caption{Overnight volatility: estimated kernel parameters for $K$'s and $L$'s, with their asymptotic confidence intervals of level $95 \%$,
                               as computed using the Fisher Information matrix with the Gaussian quantile ($1.98$).}
\label{night_coefs}
\end{table}

We define the exponential characteristic times $\tau_\text{p} := 1/\omega_\text{p}$ and $\tau_\text{e}:=1/\omega_\text{e}$, for which qualitative
interpretation is easier than for $\omega_\text{p}$ and $\omega_\text{e}$. In the case of the quadratic kernels (of type $K$), $\tau_\text{p}$ 
represents the lag where the exponential cut-off appears, after which the kernel decays to zero quickly.
One should note that in three cases, we have $\omega_\text{p}-\Delta \omega_\text{p} < 0$.
These correspond to kernels with $\alpha > 1$, which means that the power-law decays quickly by itself. In these cases 
the identifiability of $\omega_\text{p}$ is more difficult and cut-off times are ill-determined, since the value of $\omega_\text{p}$ only matters 
in a region where the kernels are already small. The exponential term $\exp(-\omega_\text{p} \, \tau)$ could be removed from the functional form of equation (\ref{eq:parametricKernels}), for these kernels only (the calibration would then modify very slightly the value of the power-law exponent $\alpha$).

\paragraph{Intra-day volatility:}

From Tab.~\ref{day_coefs} and Fig.~\ref{graph:Quadratic_kernels_DON_q512}, we see that all intra-day quadratic kernels are positive. 
However, a clear distinction is observed between intra-day feedback and overnight feedback: 
while the former is strong and decays slowly ($\alpha=0.71$ and $\tau_\text{p}=157$ days), 
      the latter decays extremely steeply ($\alpha = 2.30$) and is quickly negligible, 
      except for the intra-day immediately following the overnight, where the effect is as strong as that of the previous intra-day. 
The cross kernels ($\sstyle{ND} \to \sstyle{D}$ or $\sstyle{DN} \to \sstyle{D}$) are both statistically significant, but
are clearly smaller, and decay faster, than the $\sstyle{DD} \to \sstyle{D}$ effect. 

As far as the leverage effect is concerned, both $L$'s are found to be negative, as expected, 
and of similar decay time: $\tau_\text{e} \approx 5$ days (one week). 
However, the amplitude for their immediate impact is two times smaller for past overnight returns: 
$\LL[D]{N} = -0.0283$ versus $\LL[D]{D} = -0.0497$.

In summary, the most important part of the feedback effect on the intra-day component of the volatility 
comes from the  past intra-days themselves, except for the very last overnight, which also has a strong impact 
--- as intuitively expected, a large return overnight leads to a large immediate reaction of the market as trading resumes. 
However, this influence is seen to decay very quickly with time. 
Since a large fraction of company specific news release happen overnight, 
it is tempting to think that large overnight returns are mostly due to news. 
Our present finding would then be in line with the general result reported in \cite{joulin2008stock}: 
volatility tends to relax much faster after a news-induced jump than after endogenous jumps.

\paragraph{Overnight volatility:}

In the case of overnight volatility, Tab.~\ref{night_coefs} and Fig.~\ref{graph:Quadratic_kernels_DON_q512} illustrate
that the influence of past intra-days and past overnights is similar: 
$\KK[N]{DD}(\tau) \approx \KK[N]{NN}(\tau)$, in particular when both are large.
The cross kernels now behave quite differently:
whereas the behavior of $\KK[N]{ND}$ is not very different from that of $\KK[N]{DD}$ or $\KK[N]{NN}$ 
(although its initial amplitude is four times smaller),
$\KK[N]{DN}(\tau)$ is negative and small, but is hardly significant for $\tau \geq 2$. 
Interestingly, as pointed out above, the equality $\KK[N]{ND}=\KK[N]{DD}=\KK[N]{NN}$ means that 
it is the full Close-to-Close return that is involved in the feedback mechanism on the next overnight. 
What we find here is that this equality very roughly holds, suggesting that, as postulated in standard ARCH approaches, 
the daily close to close return is the fundamental object that feedbacks on future volatilities. 
However, this is only approximately true for the {\it overnight volatility}, 
while the intra-day volatility behaves very differently (as already said,
for intra-day returns, the largest part of the feedback mechanism comes from past intra-days only, and the very last overnight).

Finally, the leverage kernels behave very much like for the intra-day volatility. 
In fact, the $\sstyle{N} \to \sstyle{N}$ leverage kernel is very similar to its $\sstyle{N} \to \sstyle{D}$ counterpart, 
whereas the decay of the $\sstyle{D} \to \sstyle{N}$ kernel is slower ($\tau_\text{e} \approx 18$ days, nearly one month).

\paragraph{Stability and positivity:}

We checked that these empirically-determined kernels are compatible with a stable and positive volatility process.
The first obvious condition is that the system is stable with positive baseline volatilities ${\sJ}^2$. 
The self-consistent equations for the average volatilities read: 
(neglecting small cross correlations):
\begin{align}
\langle {\sigD}^2 \rangle =  {\sD}^2 	&+  \langle {\sigD}^2 \rangle \sum_{\tau=1}^\infty \KK[D]{DD}(\tau) + \langle {\sigN}^2 \rangle  \sum_{\tau=1}^\infty \KK[D]{NN}(\tau), \\    
\langle {\sigN}^2 \rangle =  {\sN}^2 	&+  \langle {\sigD}^2 \rangle \sum_{\tau=1}^\infty \KK[N]{DD}(\tau) + \langle {\sigN}^2 \rangle  \sum_{\tau=1}^\infty \KK[N]{NN}(\tau).                         
\end{align}   
This requires that the two eigenvalues of the $2\times 2$ matrix of the corresponding linear system are less than unity, i.e.
\begin{equation}\label{stability_cond}
\frac12 \left| \hK[D]{DD} + \hK[N]{NN} \pm \sqrt{ (\hK[D]{DD} - \hK[N]{NN})^2 + 4 \hK[D]{NN} \hK[N]{DD} } \right| < 1,
\end{equation}
where the hats denote the integrated kernels, schematically $\widehat{K} =\sum_{\tau=1}^\infty K(\tau)$.
This is indeed verified empirically, the eigenvalues being $\lambda_1 \simeq 0.94 \ , \ \lambda_2 \simeq 0.48$. 

Moreover, for intra-day and overnight volatilities separately, we checked that our calibrated kernels are compatible 
with the two positivity conditions derived in Appendices~\ref{appendice:nneg_vol_twoKer},\ref{appendice:nneg_vol_leverage} : 
the first one referring to the cross kernels $\KK[]{ND}$ and $\KK[]{DN}$, 
and the second one to the leverage kernels $\LL[]{D}$ and $\LL[]{N}$.
For $q=512$, the criteria fail for two spurious reasons. Firstly, for lags greater than their exponential cut-offs, the quadratic kernels vanish quicker than the `cross' kernels, which makes the \enquote{$\tau$ by $\tau$} criterion fail. Secondly, the criterion $L^\dagger K^{-1} L \leq 4 s^2$ cannot be verified for overnight volatility with ${\sN}^2=0$ (for lower values of $q$, using the same functional forms and coefficients for the kernels, ${\sN}^2$ rises to a few percents). These two effects can be considered spurious because the long-range contributions have a weak impact on the volatilities and cannot in deed generate negative values, as we checked by simulating the volatility processes with $q=512$. We thus restricted the range to $q=126$ ($=$ six months) in order to test our results with the two positive volatility criteria (again, see Appendix~\ref{appendice:nneg_vol}). For the ill-determined exponential decay rates $\omega_\text{p}$, the upper bounds of the confidence intervals are used. The two criteria are then indeed verified for both intra-day and overnight volatilities.

\subsection{Distribution of the residuals}\label{distrib_resid} 

\begin{figure}
	\centering
		\includegraphics[scale=0.6]{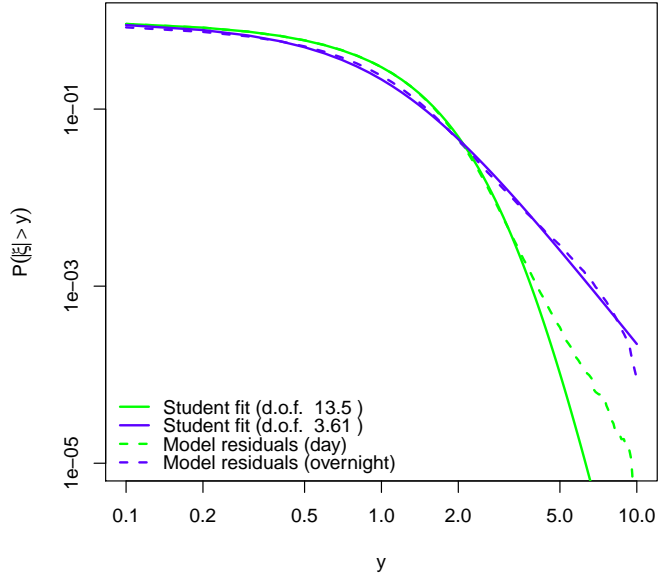}
		\caption{CDF $\mathbb{P}(|\xi| > y)$ of the residuals $\xiD$ and $\xiN$, in log-log scale.}
        \label{graph:res_distrib}
\end{figure}

As mentioned above, we assume that the residuals $\xi$ (i.e.\ the returns divided by the volatility predicted by the model) 
are Student-distributed. This is now common in ARCH/GARCH literature and was again found to be satisfactory in our companion paper 
\cite{Chicheportiche2014}. The fact that the $\xi$ are not Gaussian means that there is a residual surprise element in large
stock returns, that must be interpreted as true `jumps' in the price series.

The tail cumulative distribution function (CDF) of the residuals is shown in Fig.~\ref{graph:res_distrib}
for both intra-day and overnight returns, together with  Student best fits, obtained with long feedback kernels ($q=512$). 
This reveals a clear difference in the statistical properties of $\sss{\xi}{D}$ and $\sss{\xi}{N}$. 
First, the Student fit is  better for overnight residuals than for intra-day residuals, in particular far in the tails. 
More importantly, the number of degrees of freedom $\nu$ is markedly different for the two types of residuals:
indeed, our MLE estimation yields $\nuD=13.5$ and $\nuN=3.61$ as reported in Tab.~\ref{table:s2_nu_DON}, 
resulting in values of the residual kurtosis  $\sss{\kappa}{D} = 3 +  6/(\nuD-4) = 3.6$ and $\sss{\kappa}{N} = \infty$. 
This result must be compared with the empirical kurtosis of the returns that was measured directly in Sect.~\ref{DON_returns}. 
The intuitive conclusion is that the large (infinite?) kurtosis of the overnight returns cannot 
be attributed to fluctuations in the volatility, but rather, as mentioned above, to large jumps related to overnight news. 
This clear qualitative difference between intra-day and overnight returns is a strong argument justifying 
the need to treat these effects separately, as proposed in this paper. 

We have also studied the evolution of $\nuD$ and $\nuN$ as a function of the length $q$ of the memory of the kernels, see Tab.~\ref{table:s2_nu_DON}. 
If longer memory kernels allow to account for more of the dynamics of the volatility, less kurtic residuals should be found for larger $q$'s. 
This is indeed what we find, in particular for intra-day returns, for which $\nuD$ increases from $10.7$ for $q=21$ to $13.5$ for $q=512$. 
The increase is however much more modest for overnight returns. We propose below an interpretation of this fact.

\subsection{Baseline volatility}\label{section:baseline}

\begin{table}
\center
\begin{tabular}{|c||c|c|c|}\hline
	$q$ 			& 	21		& 	 42		&	512		\\\hline\hline
  $\nuD (\pm 0.3)$	& 	10.7 	& 	12.6 	&	13.5	\\
  $\nuN (\pm 0.02)$	& 	3.49	& 	3.58	&	3.61	\\\hline
  $\rrD$ 			& 	18.1\%	& 	12.8\%	&	8.5\%	\\
  $\rrN $			& 	14.0\%	& 	7.3\%	&	0.0\%	\\\hline
\end{tabular}
\caption{Baseline volatility and tail index of the Student residuals for several maximum lags $q$.}
\label{table:s2_nu_DON}

\end{table}

Finally, we want to study the ratio $\rrJ={{\sJ}^2}/{\langle{\sigJ}^2\rangle}$, which can be seen as a measure of the relative importance of 
the baseline component of the volatility, with respect to the endogenous feedback component.%
\footnote{In fact, the stability criterion for our model reads  $\rrJ > 0$, which is found to be 
satisfied by our calibration, albeit marginally for the overnight volatility.}
The complement $1-\rrJ$ gives the relative contribution of the feedback component, given 
by $\hK[J]{DD} + \hK[J]{NN}$ in the present context.%
\footnote{There is a contribution of the cross terms $\KK[J]{J'J''}$ since intra-day/overnight and overnight/intra-day correlations are not exactly zero,
 but this contribution is less than one order of magnitude lower than the $\hK[J]{J'J'}$.}

The results for $\rrJ$ are given in Tab.~\ref{table:s2_nu_DON} for $q=21, 42$ and $512$: as mentioned above, a larger $q$ 
explains more of the volatility, therefore reducing the value of both $\rrD$ and $\rrN$. While $\rrD$ is small ($\sim 0.1$) and comparable to 
the value found for the daily ARCH model studied in the companion paper \cite{Chicheportiche2014}, the baseline contribution is {\it nearly zero} for the 
overnight volatility. We find this result truly remarkable, and counter-intuitive at first sight. Indeed, the baseline component of the volatility is 
usually associated to exogenous factors, which, as we argued above, should be dominant for $\sigN$ since many unexpected pieces of news occur overnight!

Our interpretation of this apparent paradox relies on the highly kurtic nature of the overnight residual, with a small value $\nuN \approx 3.6$ as reported above. 
The picture is thus as follows: most overnights are news-less, in which case the overnight volatility is completely fixed by feedback effects, set by the 
influence of past returns themselves. The overnights in which important news is released, on the other hand, contribute to the tails of the residual  $\sss{\xi}{N}$, 
because the large amplitude of these returns could hardly be guessed from the previous amplitude of the returns. Furthermore, the fact that $\KK[D]{NN}$ decays 
very quickly is in agreement with the idea, expressed in \cite{joulin2008stock}, that the impact of news (chiefly concentrated overnight) on volatility is short-lived.\footnote{This effect was
confirmed recently in \cite{cosson2013master} using a direct method: the relaxation of volatility after large overnight jumps of either sign is very fast, much faster 
than the relaxation following large intra-day jumps.}

In conclusion, we find that most of the predictable part of the overnight volatility is of endogenous origin, but that the contribution of unexpected jumps 
reveals itself in the highly non-Gaussian statistics of the residuals. The intra-day volatility, on the other hand, has nearly Gaussian residuals but still a 
very large component of endogenous volatility ($1 - \rrD \approx 90\%$).

\subsection{In-Sample and Out-of-Sample tests}

In order to compare our bivariate Intra-day/Overnight volatility prediction model with the standard ARCH model for daily (close-to-close) volatility,
 we ran In-Sample (IS) and Out-of-Sample (OS) likelihood computations with both models.
Of course, in order to compare models, the same quantities must be predicted. A daily ARCH model
that predicts the daily volatility $\sigma^2_t$ at date $t$ can predict intra-day and overnight volatilities as follows:
\begin{equation}\label{eq:equivDailyVol}
{\widehat{\sigD}_t}^2 = \frac {\left[ \langle {\rD}^2 \rangle \right]} {\left[ \langle r^2 \rangle \right]} \, \sigma^2_t\ , \quad
{\widehat{\sigN}_t}^2 = \frac {\left[ \langle {\rN}^2 \rangle \right]} {\left[ \langle r^2 \rangle \right]} \, \sigma^2_t\ ,
\end{equation}
where $\left[ \langle \cdot \rangle \right]$ is the average over all dates and all stocks, 
and as in Sect.~\ref{DON_returns}, $r_t = \rD_t + \rN_t$ is the daily (close-to-close) return of date $t$.
Similarly, our bivariate intra-day/overnight model that provides predictions for intra-day and overnight volatilities separately
can give the following estimation of the daily volatility:
\begin{equation}\label{eq:equivCombinVol}
{\widehat{\sigma}_t}^2 \ = \ {\sigD_t}^2 \ + \ {\sigN_t}^2 \ + \ 2 \ [ \langle \rD \rN \rangle ].
\end{equation}
For each of the 6 predictions (of the intra-day, overnight and daily volatilities by the two models separately, bivariate Intra-day/Overnight
and standard ARCH), we use the following methodology:
\begin{itemize}
\item The pool of stock names is split in two halves,
    and the model parameters are estimated separately on each half.
\item The ``per point'' log-likelihood \eqref{eq:log_likelihood_day} is computed for both sets of parameters, 
        once with \emph{the same}  half dataset as used for the calibration (In-Sample), 
    and once with \emph{the other} half dataset (Out-of-Sample, or ``control'').
	We compute the log-likelihoods for intra-day and overnight volatilities ($\sstyle{J} \in \{ \sstyle{D},\sstyle{N} \}$), and for daily volatility:
	\begin{align*}
	\text{Bivariate models :}     \quad& \mathcal{L}^{\text{biv}} \left( \sigJ,\nuJ,\rJ \right)             \quad\text{and}\quad \mathcal{L}^{\text{biv}} \left(\widehat{\sigma},\nuDaily,\rDaily \right); \\
	\text{Standard ARCH models :} \quad& \mathcal{L}^{\text{std}} \left( \widehat{\sigJ},\nuJ,\rJ \right)   \quad\text{and}\quad \mathcal{L}^{\text{std}} \left(\sigma,\nuDaily,\rDaily \right);
	\end{align*}
	where $\mathcal{L}^{\text{biv}}$ is computed in the bivariate model (i.e.\ with six regressors: four quadratic and two linear) and $\mathcal{L}^{\text{std}}$ is
	computed in the standard ARCH model (i.e.\ with two regressors: one quadratic and one linear), and $\widehat{\sigJ}$ and $\widehat{\sigma}$
	are as given by equations (\ref{eq:equivDailyVol}) and (\ref{eq:equivCombinVol}).
	
\item The IS log-likelihood $\mathcal{L}_{\text{IS}}$ of the model is computed as the average of the two In-Sample results,
    and similarly for the OS log-likelihood $\mathcal{L}_{\text{OS}}$.
We call $l_{\text{IS}} = \exp (\mathcal{L}_{\text{IS}})$ and $l_{\text{OS}} = \exp (\mathcal{L}_{\text{OS}})$ 
the average likelihood per point (ALpp) of the model IS and OS, expressed as percentages, 
that are two proxies of the \enquote{probability that the sample data were generated by the model}.
\end{itemize}
We then use the values of $l_{\text{IS}}$ and $l_{\text{OS}}$ to compare models. For a \enquote{good} model, these values must be
as high as possible, but they must also be close to each other. As a matter of fact, if $l_{\text{IS}}$ is significantly greater
than $l_{\text{OS}}$, the model may be over-fitting the data. On the contrary, if $l_{\text{OS}}$ 
is greater than $l_{\text{IS}}$, which seems counter-intuitive, the model may be badly calibrated.
The results of this model comparison are given in Tab.~\ref{table:IS-OS_results}:
the bivariate Intra-day/Overnight model has a higher likelihood than the standard daily ARCH model,
 both In-Sample (this was to be expected even from pure over-fitting due to additional parameters)
 and Out-of-Sample (thus outperforming in predicting the ``typical'' random realization of the returns).
 
The likelihoods of the predictions obtained with equations (\ref{eq:equivDailyVol}) and (\ref{eq:equivCombinVol}) 
are marked with the exponent $\dagger$ in Tab.~\ref{table:IS-OS_results}. For these likelihoods, ``In-Sample'' simply means that the same
half of the stock pool was used for the calibration of the model and for the estimation of the likelihood, although different types 
of returns are considered. Similarly, ``Out-of-Sample'' likelihoods are estimated on the other half of the stock pool. These values
serve as comparison benchmarks between the two models.

\begin{table}[h]
\center
\begin{tabular}{|g||g|g||g|g||g|g|} \hline
	\text{Prediction}		     & 	\multicolumn{2}{g||}{{\sigD}^2} 	& 	\multicolumn{2}{g||}{{\sigN}^2} & 	\multicolumn{2}{g|}{\sigma^2 \ (\text{daily})}				\\ \hline
	\text{ALpp [\%]}		     & 	l_{\text{IS}}   & l_{\text{OS}}     & l_{\text{IS}}   &	l_{\text{OS}}   & l_{\text{IS}}		&	l_{\text{OS}}   \\ \hline \hline
  \text{Biv. Intra-day/Overnight} & 	44.423          & 44.418            & 50.839          &	50.826 			& 45.512^\dagger	&	45.509^\dagger	\\
  \text{Standard ARCH} 		     & 	44.227^\dagger  & 44.225^\dagger    & 50.598^\dagger  &	50.595^\dagger  & 44.931 			&	44.928			\\ \hline
\end{tabular}
\caption{In-Sample and Out-of-Sample average  per point likelihoods.
Figures with $\dagger$ pertain to reconstructed volatilities Eqs.~\eqref{eq:equivDailyVol} or~\eqref{eq:equivCombinVol}.}
\label{table:IS-OS_results}
\end{table}

We see that in all cases, the bivariate Intra-day/Overnight significantly outperforms the standard daily ARCH framework, in particular concerning the prediction of the total (Close-Close)
volatility. 

\section{Conclusion and extension}\label{discussion} 

The main message of this study is quite simple, and in fact to some extent expected: overnight and intra-day returns are completely different 
statistical objects. The ARCH formalism, that allows one to decompose the volatility into an exogenous component and a feedback 
component, emphasizes this difference. The salient features are:
\begin{itemize}
\item While past intra-day returns affect equally both the future intra-day and overnight volatilities, past overnight returns have a weak effect
on future intra-day volatilities (except for the very next one) but impact substantially future overnight volatilities.
\item The exogenous component of overnight volatilities is found to be close to zero, which means that the lion's share of 
overnight volatility comes from feedback effects. 
\item The residual kurtosis of returns (once the ARCH effects have been factored out) is small for intra-day returns but infinite for 
overnight returns.
\item The bivariate intra-day/overnight model significantly outperforms the standard ARCH framework based on daily returns for Out-of-Sample predictions.
\end{itemize}

Intuitively, a plausible picture for overnight returns is as follows:  most overnights are news-less, in which case the overnight volatility is completely fixed by feedback effects, set by the 
influence of past returns themselves. Some (rare) overnights witness unexpected news releases, which lead to huge jumps, the amplitude of which
could hardly have be guessed from the previous amplitude of the returns. This explains why these exogenous events contribute to residuals with such fat tails
that the kurtosis diverge, and {\it not} to the baseline volatility that concerns the majority of news-less overnights.

These conclusions hold not only for US stocks: we have performed the same study on European stocks obtaining very close model parameter estimates.%
\footnote{Equities belonging to the Bloomberg European 500 index over the same time span 2000--2009, see Appendix \ref{appendix:europe_results} for detailed results.}
Notably, the baseline volatilities are found to be $\rrD \simeq 0.1$ and $\rrN \simeq 0$ (for intra-day and overnight volatilities, respectively), 
in line with the figures found on US stocks and the interpretation drawn.
The only different qualitative behavior observed on European stocks is the quality of the Student fit for the residuals of the overnight regression:
whereas US stocks exhibit a good fit with $\nuN=3.61<4$ degrees of freedom (hence infinite kurtosis), 
European stocks have a fit of poorer quality in the tails and a parameter $\nuN=5.34$ larger than 4, hence a positive but finite kurtosis.

Having decomposed the Close-Close return into an overnight and an intra-day component, the next obvious step is to decompose the intra-day return into higher 
frequency bins --- say five minutes. We have investigated this problem as well, the results are reported in \cite{blanc2012master}. In a nutshell, we find that once the 
ARCH prediction of the intra-day average volatility is factored out, we still identify a causal feedback from past five minute returns onto the volatility of the
current bin. This feedback has again a leverage component and a quadratic (ARCH) component. The intra-day leverage kernel is close to an exponential with a 
decay time of $\approx 1$ hour. The intra-day ARCH kernel, on the other hand, {\it is still a power law}, with an exponent that we find to be close to unity, in 
agreement with several studies in the literature concerning the intra-day temporal correlations of volatility/activity --- see e.g.\ \cite{potters1998financial,liu1999statistical,wang2006scaling}, and, in the 
context of Hawkes processes, \cite{bacry2012non,hardiman2013critical}. It would be very interesting to repeat the analysis of the companion paper \cite{Chicheportiche2014} on five 
minute returns and check whether there is also a dominance of the diagonal terms of the QARCH kernels over the off-diagonal ones, as we found for daily returns. 
This would suggest a microscopic interpretation of the ARCH feedback mechanism in terms of a Hawkes process for the trading activity.

\paragraph{Acknowledgements}
We thank S.~Ciliberti, N.~Cosson, L.~Dao, C.~Emako-Kazianou, S.~Hardiman, F.~Lillo and M.~Potters for useful discussions.
P.~B. acknowledges financial support from Fondation Natixis (Paris).

\nocite{engle1986handbook}
\bibliographystyle{unsrt}
\bibliography{biblio}

\appendix

\clearpage
\section{Non-negative volatility conditions}\label{appendice:nneg_vol}

In this appendix, we study the mathematical validity of our volatility regression model.
The first obvious condition is that the model is stable, which leads to the condition (\ref{stability_cond}) in the text above.
This criterion is indeed obeyed by the kernels calibrated on empirical results.
Secondly, the volatility must remain positive, 
which is not a priori guaranteed with multiple kernels associated to signed regressors.
We now establish a set of sufficient conditions on the feedback kernels to obtain non-negative volatility processes.

\subsection{One correlation feedback kernel, no leverage coefficients}

We consider first the simpler model for daily volatility, without linear regression coefficients:
\begin{equation}
\sigma_t^2 = s^2 +   \sum_{\tau=1}^q \KK{DD}(\tau)         {\rD_{t-\tau}}^2
				 +   \sum_{\tau=1}^q \KK{NN}(\tau)         {\rN_{t-\tau}}^2
				 + 2 \sum_{\tau=1}^q \KK{ND}(\tau)   \rD_{t-\tau} \rN_{t-\tau} \nonumber.
\end{equation}
This modification of the standard ARCH model can lead to negative volatilities if 
(at least) one term in the last sum 
takes large negative values. 
This issue can be studied more precisely with the matrix form of the model:
\[ \sigma_t^2 = s^2 + R_t^\dagger K R_t, \]
with
$$
  K = 
 \begin{pmatrix}
\KK{DD}(1)	&	    	&			&\KK{ND}(1)	&	    	&	 		\\
            &	\ddots	&			&	    	&	\ddots	&	 		\\
		    &			&\KK{DD}(q)	&			&			&\KK{ND}(q) 	\\
\KK{ND}(1)	&	    	&			&\KK{NN}(1)	&	    	&	 		\\
            &	\ddots	&			&	    	&	\ddots	&	 		\\
		    &			&\KK{ND}(q)	&			&			&\KK{NN}(q) 	\\
\end{pmatrix}
\ ,\quad
R_t = 
\begin{pmatrix}
\rD_{t-1} \\
\vdots \\
\rD_{t-q} \\
\rN_{t-1} \\
\vdots \\
\rN_{t-q} \\
\end{pmatrix},
$$
and where $\KK{DD}$ and $\KK{NN}$ coefficients are assumed to be all positive (which is the case empirically).
This formula highlights the fact that the volatility remains positive 
as soon as the symmetric matrix $K$ is positive semidefinite. 
We now determine a sufficient and necessary condition under which $K$ has negative eigenvalues. 
The characteristic polynomial of $K$ is
\[
\chi_K (X) = \prod_{\tau = 1}^q \Big{[}(\KK{DD}(\tau)-X)(\KK{NN}(\tau)-X)-\KK{ND}(\tau)^2 \Big{]},
\]
and the eigenvalues of $K$ are the zeros of $\chi_K (X)$, solutions $\chi_K(\lambda)=0$, 
i.e.\ such that
$$
\lambda^2 - (\KK{DD}(\tau)+\KK{NN}(\tau)) \lambda + \KK{DD}(\tau)\KK{NN}(\tau) -\KK{ND}(\tau)^2 = 0.
$$
Hence, $K$ has at least one negative eigenvalue iff 
$\exists \tau \in \{ 1, \ldots , q \}$ s.t.
\[ 
    \KK{DD}(\tau)+\KK{NN}(\tau)-\sqrt{(\KK{DD}(\tau)-\KK{NN}(\tau))^2 + 4 \KK{ND}(\tau)^2} < 0,
\]
and finally,
\begin{equation}\label{eq:nnegvol}
K \ \text{is positive semidefinite} \
\Leftrightarrow \
\forall \tau \in \{ 1, \ldots , q \}, \ \ \KK{ND}(\tau)^2 \leq \KK{DD}(\tau) \KK{NN}(\tau).
\end{equation}
When the quadratic kernel $K$ is positive-semidefinite, $\sigma_t^2$ remains positive for all $t$.
For example, in the standard ARCH model, the inequality is saturated for all $\tau$ by construction, and the condition~\eqref{eq:nnegvol} is satisfied.

The positive-semidefiniteness of $K$, equivalent to 
$$ \forall v = \left(\vD_1, \ldots, \vD_q, \vN_1, \ldots, \vN_q \right)^\dagger \in \mathbb{R}^{2q} \ , \quad v^\dagger K v \geq 0,$$
is ensured by the \emph{sufficient} condition that every term in the development of the quadratic form is positive:
\begin{align*}  
K & \text{ is positive-semidefinite} \\
  &\Leftarrow \forall v, \ \forall \tau \in \{1,\ldots,q\}, \quad 
                 \KK{DD}(\tau) {\vD_\tau}^2 + 
                 \KK{NN}(\tau) {\vN_\tau}^2 + 
               2 \KK{ND}(\tau) \vD_\tau \vN_\tau  \geq 0 	                         \\
  &\Leftarrow \forall v, \ \forall \tau \in \{1,\ldots,q\}, \quad 
                | \KK{ND}(\tau) \vD_\tau \vN_\tau | \leq 
                  \frac12 (\KK{DD}(\tau) {\vD_\tau}^2 +  \KK{NN}(\tau) {\vN_\tau}^2)	 \\
  &\Leftarrow \forall \tau \in \{1,\ldots,q\}, \quad 
                |\KK{ND}(\tau)| \leq \sqrt{\KK{DD}(\tau) \KK{NN}(\tau)}                   
\end{align*}
Although more stringent {\it a priori}, this \enquote{$\tau$ by $\tau$} condition
resumes, in this particular case, to the necessary \emph{and} sufficient criterion~\eqref{eq:nnegvol}.
In the next subsection, we use the same method to obtain a sufficient condition for the semidefiniteness of $K$
in the more complicated case with two correlation feedback kernels.

\subsection{Two correlation feedback kernels, no leverage coefficients}\label{appendice:nneg_vol_twoKer}
With an additional feedback function $\KK{DN}$
and a coefficient $\KK{NN}(0)$ corresponding to the $\tau = 0$ term in the ${\rN}^2$ sum,
the model is
\begin{align*}
\sigma_t^2 = s^2 &+   \sum_{\tau=1}^q \KK{DD}(\tau) {\rD_{t-\tau}}^2
				  +   \sum_{\tau=0}^q \KK{NN}(\tau) {\rN_{t-\tau}}^2\\
				 &+ 2 \sum_{\tau=1}^q \KK{ND}(\tau) \rD_{t-\tau} \rN_{t-\tau}
				  + 2 \sum_{\tau=1}^q \KK{DN}(\tau) \rD_{t-\tau} \rN_{t-\tau+1},
\end{align*}
or $\sigma_t^2 = s^2 + R_t^\dagger K R_t$,
with $K \in \mathbb{R}^{(2q+1) \times (2q+1)}$, $R_t \in \mathbb{R}^{(2q+1)}$ defined by
$$
 K = 
 \begin{pmatrix}
\KK{DD}(1)	&	    	&				&	\KK{DN}(1)		&	\KK{ND}(1)	&	     		&				\\
            &	\ddots	&				&	    			&	\ddots		&	\ddots		&				\\
			&			&	\KK{DD}(q)	&					&	     		&	\KK{DN}(q)	&	\KK{ND}(q)	\\
\KK{DN}(1)	&	    	&				&	\KK{NN}(0)		&				&	     		&	 			\\
\KK{ND}(1)	&	\ddots	&	    		&					&	\KK{NN}(1) 	&				&	     		\\
            &	\ddots	&	\KK{DN}(q)	&	    			&				&	\ddots		&			 	\\
			&			&	\KK{ND}(q)	&					&	     		&				&	\KK{NN}(q) 	\\
\end{pmatrix},\quad
R_t =
\begin{pmatrix}
\rD_{t-1} \\
\vdots \\
\rD_{t-q} \\
\rN_t \\
\rN_{t-1} \\
\vdots \\
\rN_{t-q} \\
\end{pmatrix}.
$$
The positive-semidefiniteness of $K$ is harder to characterize directly, 
so we use the \enquote{$\tau$ by $\tau$} method to find a criterion for a sufficient condition.
For any $\beta \in ]0,1[$ and any vector 
$
v = \left(\vD_1, \ldots, \vD_q, \vN_0, \vN_1, \ldots, \vN_q \right)^\dagger
\in \mathbb{R}^{(2q+1)}
$, 
the quadratic form $v^\dagger K v$ is decomposed as follows:
\begin{align*}
v^\dagger K v 	  = & \KK{NN}(0) {\vN_0}^2  +  \sum_{\tau=1}^q \Big[ \KK{DD}(\tau) {\vD_\tau}^2 
                    +   \KK{NN}(\tau) {\vN_\tau}^2 
                    + 2 \KK{ND}(\tau) \vD_\tau \vN_\tau 
                    + 2 \KK{DN}(\tau) \vD_\tau \vN_{\tau-1} \Big]\\
                  = & \phantom{+\sum_{\tau=1}^{q} \Big[}
                        \beta \KK{NN}(0) {\vN_0}^2              +   (1-\beta) \KK{NN}(q) {\vN_q}^2\\  
                   &+ \sum_{\tau=1}^{q} \Big[ 
                        \beta \KK{NN}(\tau  ) {\vN_{\tau  }}^2  +   (1-\beta) \KK{NN}(\tau-1) {\vN_{\tau-1}}^2  \\  
                   &\phantom{+\sum_{\tau=1}^{q} \Big[}
                    +   \KK{DD}(\tau) {\vD_\tau}^2 
                    + 2 \KK{ND}(\tau) \vD_\tau \vN_{\tau  }
                    + 2 \KK{DN}(\tau) \vD_\tau \vN_{\tau-1} \Big],
\end{align*}
and clearly, a sufficient condition for the sum to be non-negative is that each term is non-negative:
\begin{align*}
\forall t, \ \sigma_t^2 \geq 0
\Leftarrow & \ K \ \text{is positive-semidefinite}\\
\Leftarrow & \quad \exists \beta \in ]0,1[ \ , \ \forall v = \left(\vD_1, \ldots, \vD_q, \vN_0, \vN_1, \ldots, \vN_q \right)^\dagger
\in \mathbb{R}^{(2q+1)},
\forall \tau \in \{1,\ldots,q\},\\
&\quad 0\leq \beta      \KK{NN}(\tau) {\vN_\tau}^2 + (1-\beta) \KK{NN}(\tau-1) {\vN_{\tau-1}}^2 \\
&\phantom{\quad 0 } +\KK{DD}(\tau) {\vD_\tau}^2 + 2 \KK{ND}(\tau) \vD_\tau \vN_\tau  + 2 \KK{DN}(\tau) \vD_\tau \vN_{\tau-1}\\
\Leftarrow  & \quad \exists \beta, \ \forall v, \ \forall \tau \in \{1,\ldots,q\}, \quad \exists \alpha_{\tau} \in [0,1],\\
\bullet	    & \quad	| \KK{ND}(\tau) \vD_\tau \vN_\tau | \leq
						\frac12 \Big( \alpha_{\tau} \ \KK{DD}(\tau) {\vD_\tau}^2 + \beta \KK{NN}(\tau) {\vN_\tau}^2 \Big)\\
\bullet     & \quad	| \KK{DN}(\tau) \vD_\tau \vN_{\tau-1} | \leq
						\frac12 \Big( (1-\alpha_{\tau}) \ \KK{DD}(\tau) {\vD_\tau}^2 + (1-\beta) \KK{NN}(\tau-1) {\vN_{\tau-1}}^2 \Big)\\
\Leftarrow  & \quad \exists \beta, \ \forall \tau \in \{1,\ldots,q\}, \quad \exists \alpha_{\tau} \in [0,1],\\
\bullet		& \quad	\KK{ND}(\tau)^2 \leq \beta \ \alpha_{\tau} \ \KK{DD}(\tau) \KK{NN}(\tau)\\
\bullet		& \quad	\KK{DN}(\tau)^2 \leq (1-\beta) (1-\alpha_{\tau}) \ \KK{DD}(\tau) \KK{NN}(\tau-1).
\end{align*}
The last condition is equivalent to a simpler one, with $\alpha_{\tau}$ saturating one of the two inequalities:
denoting $\sss{\delta}{(nn)}(\tau) = {\KK{NN}(\tau)}/{\KK{NN}(\tau-1)}$
and      $\sss{\delta}{(nd)}(\tau) = {\KK{DN}(\tau)}/{\KK{ND}(\tau)}$,
$K$ is positive-semidefinite if (but not only if), $\exists \beta \in ]0,1[, \ \forall \tau \in \{1,\ldots,q\}$, 
\begin{align*}
	\mathcal{M}(\beta,\tau)\equiv\max \Bigg\{&
                         \frac {\beta \sss{\delta}{(nn)}(\tau) } {1-\beta} \left(\frac{(1-\beta) \KK{DD}(\tau) \KK{NN}(\tau-1)}{\KK{ND}(\tau)^2} -           \sss{\delta}{(nd)}(\tau)^2 \right) \ ,\\
			  & \frac {1-\beta} {\beta \sss{\delta}{(nn)}(\tau) } \left(\frac{\beta \KK{DD}(\tau) \KK{NN}(\tau  )}{\KK{DN}(\tau)^2} -  \frac{1}{\sss{\delta}{(nd)}(\tau)^2}\right)
			  \Bigg\}
\end{align*}
is larger than one, yielding the following a.s.\ positive volatility criterion:
$$\forall t, \ \sigma_t^2 \geq 0 \Leftarrow 1\leq \sup_{\beta \in ]0,1[} \min_{1 \leq \tau \leq q}\mathcal{M}(\beta,\tau).$$

\subsection{With leverage coefficients}\label{appendice:nneg_vol_leverage}
We now add leverage terms to the volatility equation:
$$ \sigma_t^2 = s^2 + R_t^\dagger K R_t + R_t^\dagger L 
  = \widetilde{R}_t^\dagger M \widetilde{R}_t,$$
with
$$ M = 
 \begin{pmatrix}
	 K				         		&	\frac12 {L}	\\
 	 \frac12 {L^\dagger}			&	s^2			\\	
\end{pmatrix},
$$
and appropriate vectors $R_t, L_t,\widetilde{R}_t$.
It is easy to show that, assuming a positive-definite $K$,
\begin{equation}\label{eq:nnegvol_lev}
M \ \text{is positive-semidefinite}
\ \Leftrightarrow \ L^\dagger K^{-1} L \leq 4 s^2.
\end{equation}

\clearpage
\section{Universality assumption}\label{appendix:univ_DON} 

To obtain a better convergence of the parameters of the model, the estimates are averaged over a pool of $280$ US stocks. 
The validity of this method lies on the assumption that the model is approximately universal, 
i.e.\ that the values of its coefficients do not vary significantly among stocks.

We check that this assumption is relevant by splitting the stock pool in two halves 
and running the estimation of the model on the two halves independently. We obtain a set $\Theta_1 \in \mathbb{R}^{17}$
of coefficients calibrated on the first half and a set $\Theta_2 \in \mathbb{R}^{17}$ on the second half (each set contains $17$ parameters,
three for each of the four $K$ kernels, two for each of the two $L$ kernels, plus $\nu$).

If the (normalized) returns series for each stock were realizations of the same process, the differences between the coefficients
of the two half stock pools would be explained by statistical noise only. To quantify how close the observed differences are to statistical noise,
we run a series of Wald tests and study the obtained p-values. We test $H_0 = \{ f(\Theta) = 0 \}$ against $H_1 = \{ f(\Theta) \neq 0 \}$, where
$\Theta = (\Theta_1,\Theta_2) \in \mathbb{R}^{34} \ , \ f(\Theta) = \Theta_1 - \Theta_2 \ , \ f : \mathbb{R}^{34} \mapsto \mathbb{R}^{17}$,
by comparing the statistic
\begin{equation}
\Xi_n =  n \, f(\Theta)^\dagger \, \Sigma(\Theta)^{-1} \, f(\Theta) 
, \quad\text{with}\quad 
\Sigma(\Theta) = \frac{\partial f}{\partial \Theta}(\Theta) \, I(\Theta)^{-1} \, \left( \frac {\partial f} {\partial \Theta} (\Theta) \right)^\dagger,
\end{equation}
to the quantiles of a $\chi^2$ variable, where $n=\frac12 \times 280 \times 2515$ is the sample size for each half stock pool,
$I(\Theta)$ is the Fisher Information matrix of the model and $\frac {\partial f} {\partial \Theta}$ is the Jacobian matrix of $f(\Theta)$.

For intra-day volatility, the p-value is close to zero if all the $17$ coefficients are included, but becomes very high ($\text{p-val} = 12.3 \%$) if we
exclude $\alpha^{\sstyle{NN} \to \sstyle{D}}$ from the test. One can conclude that all the parameters but $\alpha^{\sstyle{NN} \to \sstyle{D}}$
can be considered universal, with a high significance level of $10 \%$. It is not surprising that at least one coefficient varies slightly among stocks
(it would indeed be a huge discovery to find that $280$ US stocks can be considered as identically distributed!).

In the case of overnight volatility, we first test the universality of the parameters in $\KK[N]{DD}$ and $\KK[N]{NN}$, 
for which the constraint ${\sN}^2 = 0$ is included in the estimation. 
We then test the other $11$ parameters for universality. Once again, a few of them
($g_\text{p}^{\sstyle{ND} \to \sstyle{N}}$, $\alpha^{\sstyle{NN} \to \sstyle{N}}$
and $\omega^{\sstyle{NN} \to \sstyle{N}}_\text{p}$) must be excluded from
the tests to obtain acceptable p-values. We then obtain $\text{p-val} = 1 \%$ for the first test and $\text{p-val} = 7.5 \%$ for the second.

It is then natural to wonder whether the four coefficients that cannot be considered as statistically universal differ significantly in 
relative values.
That is why we compute a second comparison criterion: for a pair $(c^{(1)},c^{(2)})$
of coefficients estimated on the first and second half stock pools respectively, we compute the relative
difference, defined as:
\[
\quad \frac{|c^{(1)}-c^{(2)}|}{\max\{|c^{(1)}|,|c^{(2)}|\}}.
\]
The values of this criterion are summarized in Tabs.~\ref{day_univ_table_relative_diff},\ref{night_univ_table_relative_diff}.
The first observation is that no relative difference exceeds $50 \%$ (except for three of the ill-determined
$\omega_\text{p}$) which indicates that the signs and orders of magnitude
of the coefficients of the model are invariant among stocks.
The coefficients for which the relative difference is high but the statistical one is low do not contradict the universality assumption:
the ML estimation would need a larger dataset to determine them with precision, and the difference between the two halves can be 
interpreted as statistical noise.

Three of the four \enquote{non-universal} coefficients, $\alpha^{\sstyle{NN} \to \sstyle{D}}$, $g_\text{p}^{\sstyle{ND} \to \sstyle{N}}$
and $\omega^{\sstyle{NN} \to \sstyle{N}}_\text{p}$ also show a significant relative difference between the two stock pools (above $10\%$).
These are thus the only coefficients of the model for which averaging over all stocks is in principle not well-justified, and for which the confidence intervals given
in Sect.~\ref{section:kernels} should be larger. However, these variations do not impact the global shapes
of the corresponding kernels in a major way, and our qualitative comments on the feedback of past returns
on future intra-day and overnight volatilities are still valid.

The results of this section indicate that most coefficients of the model are compatible with the assumption of universality.
Although some coefficients do show slight variations, our stock aggregation method
(with proper normalization, as presented in Sect.~\ref{subsection:data}) is reasonable.

\begin{table}[h]
\center
\begin{tabular}{|g||g|g|g||} \hline
	\text{Kernels}	& 	\multicolumn{3}{g||}{K(\tau)=g_\text{p} \, \tau^{-\alpha} \, e^{-\omega_\text{p} \, \tau}    } 							\\\hline
					& 	g_\text{p}						& \alpha 									& \omega_\text{p} 							\\\hline\hline
  \KK[D]{DD} 		& 	\textcolor{darkgreen}{9.4 \%} 	& \textcolor{darkgreen}{7.3 \%}				& \textcolor{darkgreen}{13.1 \%}				\\
  \KK[D]{NN} 		& 	\textcolor{darkgreen}{2.6 \%} 	& \textcolor{darkgreen}{17.2 \%}	& \textcolor{darkgreen}{\textbf{34.4 \%}}	\\
  \KK[D]{ND} 		& 	\textcolor{darkgreen}{13.3 \%}	& \textcolor{darkgreen}{9.6 \%}				& \textcolor{darkgreen}{\textbf{77.6 \%}}				\\
  \KK[D]{DN} 		& 	\textcolor{darkgreen}{2.0 \%} 	& \textcolor{darkgreen}{9.9 \%}				& \textcolor{darkgreen}{\textbf{94.8 \%}}	\\\hline
\end{tabular}
\begin{tabular}{|g||g|g||} \hline
	\text{Kernels}	& 	\multicolumn{2}{g||}{L(\tau)=g_\text{e} \, e^{-\omega_\text{e} \, \tau}   }  			\\\hline
					&	g_\text{e} 									& \omega_\text{e} 					\\\hline\hline
  \LL[D]{D} 			&	\textcolor{darkgreen}{10.4 \%}		& \textcolor{darkgreen}{1.8 \%}		\\ && \\
  \LL[D]{N} 			&	\textcolor{darkgreen}{5.8 \%}				& \textcolor{darkgreen}{2.3 \%}		\\ && \\\hline
\end{tabular}
\caption{Intra-day volatility: relative differences between the two half stock pools ($q=512$). 
         For $\nuD$, the value is $7.1 \%$.
         Bold figures are above $20 \%$.}
\label{day_univ_table_relative_diff}

\end{table}

\begin{table}[h]
\center

\begin{tabular}{|g||g|g|g||} \hline
	\text{Kernels}	& 	\multicolumn{3}{g||}{K(\tau)=g_\text{p} \, \tau^{-\alpha} \, e^{-\omega_\text{p} \, \tau}       } 		\\\hline
					& 	g_\text{p}					& \alpha 								& \omega_\text{p} 				\\\hline\hline
  \KK[N]{DD} 		& 	\textcolor{blue}{3.1 \%} 		& \textcolor{blue}{9.7 \%}				& \textcolor{blue}{17.0 \%}		\\
  \KK[N]{NN} 		& 	\textcolor{blue}{8.3 \%} 		& \textcolor{blue}{7.6 \%}				& \textcolor{blue}{\textbf{33.0 \%}}		\\
  \KK[N]{ND} 		& 	\textcolor{blue}{\textbf{30.1 \%}}		& \textcolor{blue}{1.7 \%}				& \textcolor{blue}{\textbf{80.0 \%}}		\\
  \KK[N]{DN} 		& 	\textcolor{blue}{\textbf{35.5 \%}} 	& \textcolor{blue}{\textbf{21.0 \%}}	& \textcolor{blue}{0.2 \%}		\\\hline
\end{tabular}
\begin{tabular}{|g||g|g||} \hline
	\text{Kernels}	& 	\multicolumn{2}{g||}{L(\tau)=g_\text{e} \, e^{-\omega_\text{e} \, \tau}   }  			\\\hline
					&	g_\text{e}								& \omega_\text{e} 						\\\hline\hline
  \LL[N]{D} 			&	\textcolor{blue}{18.3 \%}		& \textcolor{blue}{\textbf{32.0 \%}}	\\ && \\
  \LL[N]{N} 			&	\textcolor{blue}{19.4 \%}		& \textcolor{blue}{6.5 \%}				\\ && \\\hline
\end{tabular}
\caption{Overnight volatility: relative differences between the two half stock pools ($q=512$). 
        For $\nuN$ the value is $0.03 \%$.
        Bold figures are above $20 \%$.}
\label{night_univ_table_relative_diff}

\end{table}

 \clearpage
\section{The case of European stocks: results and discussions}\label{appendix:europe_results}

In order to verify that our conclusions are global and not specific to US stock markets,
 we also calibrated our model on European stock returns. 
 The dataset is composed of daily prices for
$179$ European stocks of the Bloomberg European 500 index, on the same period 2000--2009. 
The data treatment is exactly the same as before.
The following sections analyze and compare the results to those obtained on US stocks.

\subsection{The feedback kernels: parameters estimates}
\begin{figure}[p]
	\centering
    \subfigure[Quadratic $K(\tau)$ kernels in log-log scale. For `DN$\to$N', the absolute value of the (negative) kernel is plotted.]{
    \includegraphics[scale=0.5]{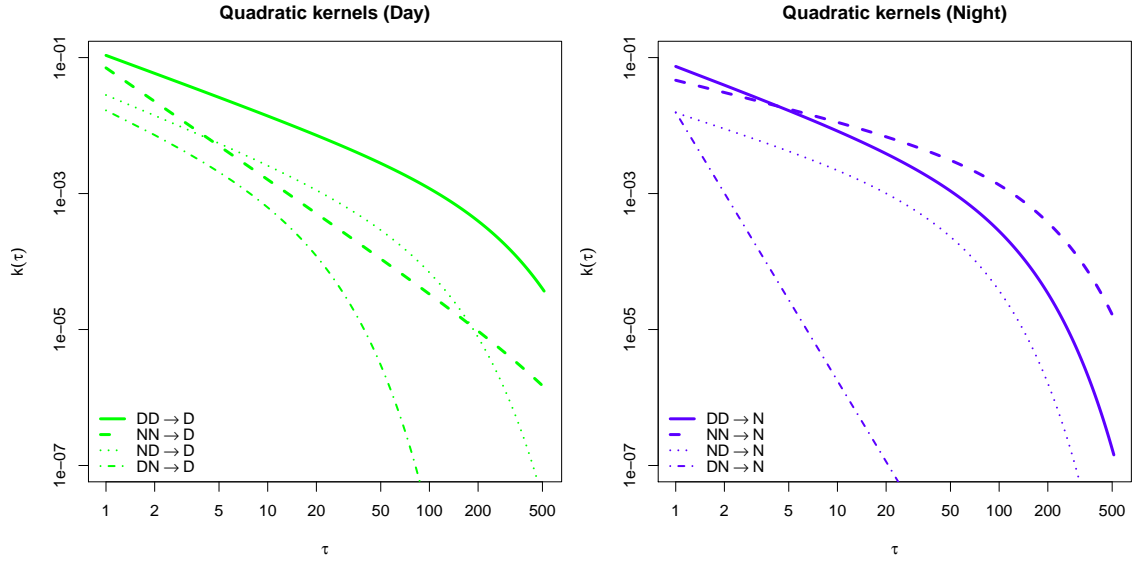}
	\label{graph:euro_Quadratic_kernels_DON_q512}
    }
	\subfigure[Linear $L(\tau)$ kernels in lin-log scale. All leverage kernels are negative, so $-L(\tau)$ are plotted.]{
    \includegraphics[scale=0.6]{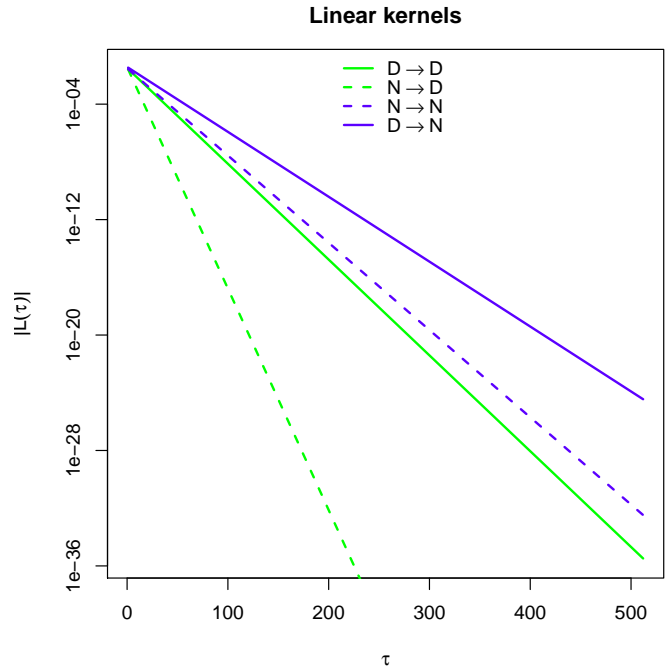}
	\label{graph:euro_Leverage_kernels_DON_q512} 
    }
	\caption{Estimated kernels, impacting Intra-Days in (light) green, OverNights in (dark) blue.}
\end{figure}

ML estimates of the parameters in the regression kernels (for a maximum lag $q=512$) are reported in Tabs.~\ref{euro_day_coefs},\ref{euro_night_coefs}, 
and the resulting kernels are shown in Figs.~\ref{graph:euro_Quadratic_kernels_DON_q512},\ref{graph:euro_Leverage_kernels_DON_q512}.

\begin{table}[h]
\center
\begin{tabular}{|g||g|g|g||g|g|} \hline
	\text{Kernels}	& 	\multicolumn{3}{g||}{K(\tau)=g_\text{p} \, \tau^{-\alpha} \, e^{-\omega_\text{p} \times \tau}    } 									& 	\multicolumn{2}{g||}{L(\tau)=g_\text{e} \, e^{-\omega_\text{e} \times \tau}}  			\\\hline
					& 	g_\text{p} \times 10^2					& \alpha 									& \omega_\text{p} \times 10^2					&	g_\text{e} \times 10^2					& \omega_\text{e} \times 10^2					\\\hline\hline
  \KK[D]{DD} 		& 	\textcolor{darkgreen}{10.83 \pm 0.11} 	& \textcolor{darkgreen}{0.87 \pm 0.005}		& \textcolor{darkgreen}{0.50 \pm 0.03}			&	--										& --											\\
  \KK[D]{NN} 		& 	\textcolor{darkgreen}{7.06 \pm 0.24} 	& \textcolor{darkgreen}{1.64 \pm 0.03}		& \textcolor{darkgreen}{0.12 \pm 0.28}			&	--										& --											\\
  \KK[D]{ND} 		& 	\textcolor{darkgreen}{2.86 \pm 0.19}	& \textcolor{darkgreen}{0.98 \pm 0.06}		& \textcolor{darkgreen}{1.51 \pm 0.84}			&	--	    								& --											\\
  \KK[D]{DN} 		& 	\textcolor{darkgreen}{1.83 \pm 0.30} 	& \textcolor{darkgreen}{1.08 \pm 0.26}		& \textcolor{darkgreen}{9.02 \pm 8.11}			&	--										& --											\\
  \LL[D]{D} 			& 	--										& --		    							& --											&	\textcolor{darkgreen}{-3.20 \pm 0.27}	& \textcolor{darkgreen}{15.29 \pm 1.39}			\\
  \LL[D]{N} 			& 	--										& --		   								& --											&	\textcolor{darkgreen}{-3.73 \pm 0.51}	& \textcolor{darkgreen}{35.35 \pm 4.69}			\\\hline
\end{tabular}
\caption{Intra-day volatility: estimated kernel parameters for $K$'s and $L$'s, with their asymptotic confidence intervals of level $95 \%$,
as computed using the Fisher Information matrix with the Gaussian quantile ($1.98$).}
\label{euro_day_coefs}

\center
\begin{tabular}{|g||g|g|g||g|g|} \hline
	\text{Kernels}	& 	\multicolumn{3}{g||}{K(\tau)=g_\text{p} \, \tau^{-\alpha} \, e^{-\omega_\text{p} \times \tau}    } 			& 	\multicolumn{2}{g||}{L(\tau)=g_\text{e} \, e^{-\omega_\text{e} \times \tau}}  	\\\hline
					& 	g_\text{p} \times 10^2 			& \alpha 							& \omega_\text{p} \times 10^2		&	g_\text{e} \times 10^2  			& \omega_\text{e} \times 10^2	 			\\\hline\hline
  \KK[N]{DD}		& 	\textcolor{blue}{7.53 \pm 0.39} 	& \textcolor{blue}{0.89 \pm 0.03}	& \textcolor{blue}{1.49 \pm 0.57}	&	--								& --							    		\\
  \KK[N]{NN}		& 	\textcolor{blue}{4.69 \pm 0.25} 	& \textcolor{blue}{0.58 \pm 0.01}	& \textcolor{blue}{0.86 \pm 0.07}	&	--								& --						        		\\
  \KK[N]{ND}   		& 	\textcolor{blue}{1.59 \pm 0.20}	& \textcolor{blue}{0.75 \pm 0.14}	& \textcolor{blue}{2.62 \pm 2.07}	&	--   								& --						        		\\
  \KK[N]{DN}   		& 	\textcolor{blue}{-1.57 \pm 0.33} 	& \textcolor{blue}{3.95 \pm 1.32}	& \textcolor{blue}{0.02 \pm 13.67}	&	--								& --				  					\\
  \LL[N]{D} 			& 	--								& --	    								& --									&	\textcolor{blue}{-3.78 \pm 0.23}	& \textcolor{blue}{10.36 \pm 0.71}	\\
  \LL[N]{N}  			& 	--								& --	    								& --						    			&	\textcolor{blue}{-3.09 \pm 0.29}	& \textcolor{blue}{13.93 \pm 1.74}	\\\hline
\end{tabular}
\caption{Overnight volatility: estimated kernel parameters for $K$'s and $L$'s, with their asymptotic confidence intervals of level $95 \%$,
as computed using the Fisher Information matrix with the Gaussian quantile ($1.98$).}
\label{euro_night_coefs}
\end{table}

\paragraph{Intra-day volatility:}

From Tab.~\ref{euro_day_coefs} and Fig.~\ref{graph:euro_Quadratic_kernels_DON_q512}, we see that all the conclusions drawn previously for the case of US stocks hold
for European stocks. The intra-day feedback is stronger and of much longer memory than overnight feedback, which decays very quickly (although
more slowly for European stocks, with $\alpha \simeq 1.6$ instead of $\alpha \simeq 2.3$). 
The cross kernels are still clearly smaller than the two quadratic ones, with $\alpha$ close to unity.

The leverage effect is similar to the case of US stocks too, although its initial amplitude is approximately equal
for past intra-day and overnight returns, whereas past intra-days are stronger than overnights for US stocks.

\paragraph{Overnight volatility:}

In the case of overnight volatility, we see from Tab.~\ref{euro_night_coefs} and Fig.~\ref{graph:euro_Quadratic_kernels_DON_q512} that 
not only do all our previous conclusions still hold in the European case (long memory 
for both intra-day and overnight feedback, second cross kernel nearly equal to zero), but the
coefficients of the model are remarkably close to those of the US calibration.

\subsection{Distribution of the residuals}

\begin{figure}
	\centering
		\includegraphics[scale=0.6]{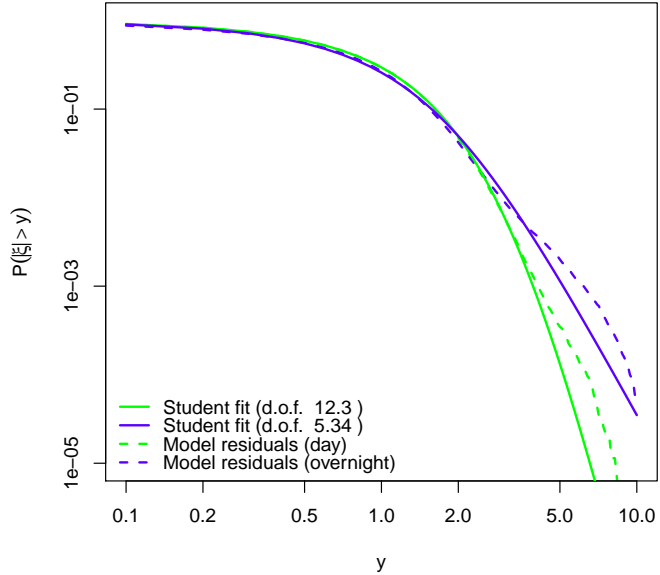}
		\caption{CDF $\mathbb{P}(|\xi| > y)$ of the residuals $\xiD$ and $\xiN$, in log-log scale.}
        \label{graph:euro_res_distrib}
\end{figure}

For intra-day returns, the distribution of the residuals is very similar to the case of US stocks.
However, for overnight returns, some differences must be pointed out. Firstly, as can be seen on Tab.~\ref{table:euro_s2_nu_DON},
 $\nuN$ is significantly higher for European stock ($5.3$) than for US stocks ($3.6$). 
 As a consequence, the kurtosis of overnight residuals is finite: $3+\frac6{\nuN-4} \simeq 7.5$.
 The distribution is still highly leptokurtic, but the result is less extreme than for US stocks. 
 Secondly, figure \ref{graph:euro_res_distrib} shows that the quality of the Student fit is of lesser quality here. 
 For European stocks, both intra-day and overnight residuals seem to be fitted by a lower value of $\nu$ for far tail events, 
 whereas this only held for intra-day residuals in the US case.

\subsection{Baseline volatility}

\begin{table}
\center
\begin{tabular}{|c||c|}\hline
	$q$ 			&	512		\\\hline\hline
  $\nuD (\pm 0.4)$	&	12.3	\\
  $\nuN (\pm 0.06)$	&	5.34	\\\hline
  $\rrD$ 			&	10.0\%	\\
  $\rrN $			&	0.0\%	\\\hline
\end{tabular}
\caption{Baseline volatility and tail index of the Student residuals for $q = 512$.}
\label{table:euro_s2_nu_DON}

\end{table}

Finally, we compare the ratios $\rrJ={{\sJ}^2}/{\langle{\sigJ}^2\rangle}$ of the two stock pools. 
Here again, the results are very close to our previous calibration: $\rrD \simeq 0.1$ for intra-day volatility, $\rrN \simeq 0$ for overnight volatility. 
Like in the case of US stocks, the calibration procedure yields a slightly negative ${\sN}^2$, 
so we add an additional step that includes the constraint ${\sN}^2 = 0$ (for overnight volatility only). 

One of our main conclusions for US stocks is that overnight volatility is entirely endogenous, 
and that the exogeity of overnight returns is contained in the leptokurtic distribution of overnight residuals. 
This section proves that this is also true for European stocks and suggests that our findings hold quite generally.

 \end{document}